\documentclass[fleqn,usenatbib]{mnras}

\usepackage{newtxtext,newtxmath}
\usepackage[T1]{fontenc}

\DeclareRobustCommand{\VAN}[3]{#2}
\let\VANthebibliography\thebibliography
\def\thebibliography{\DeclareRobustCommand{\VAN}[3]{##3}\VANthebibliography}

\usepackage{graphicx}	
\usepackage{amsmath}

\title[Starburst Fragmentation]{The Fragmentation of Molecular Clouds in Starburst Environments}

\author[M. T. Cusack et al.]{
Matt T. Cusack,$^{1}$\thanks{E-mail: cusackmt@cardiff.ac.uk}
Paul C. Clark,$^{1}$
Simon C. O. Glover,$^{2}$
Ralf S. Klessen,$^{2,3,4,5}$
Philipp Girichidis,$^{2}$
\newauthor{
Anthony P. Whitworth,$^{1}$ 
Felix D. Priestley$^{1}$}
\\
$^{1}$School of Physics and Astronomy, Cardiff University, Queen’s Buildings, The Parade, Cardiff CF24 3AA, UK  \\
$^{2}$Universit\"{a}t Heidelberg, Zentrum für Astronomie, Institut für Theortische Astrophysik, Albert-Ueberle-Str. 2, 69120 Heidelberg, Germany \\
$^{3}$Universit\"{a}t Heidelberg, Interdisziplin\"{a}res Zentrum f\"{u}r Wissenschaftliches Rechnen, Im Neuenheimer Feld 225, 69120 Heidelberg, Germany \label{IWR} \\
$^{4}$Harvard-Smithsonian Center for Astrophysics, 60 Garden Street, Cambridge, MA 02138, U.S.A. \label{CfA} \\
$^{5}$Elizabeth S. and Richard M. Cashin Fellow at the Radcliffe Institute for Advanced Studies at Harvard University, 10 Garden Street, Cambridge, MA 02138, U.S.A. \label{Radcliffe} }

\pubyear{2024}

\begin{document}
\label{firstpage}
\pagerange{\pageref{firstpage}--\pageref{lastpage}}
\maketitle

% Removing the headers
\thispagestyle{plain}

%%%%%%%%%%%%%%%%%%% ABSTRACT %%%%%%%%%%%%%%%%%%%%%

\begin{abstract}
    A significant amount of star formation occurs and has occurred in environments unlike the solar neighbourhood. The majority of stars formed closer to the peak of the cosmic star formation rate ($z \rm > 1.3$) and a great deal of star formation presently occurs in the central molecular zone (CMZ) of the Galaxy. These environments are unified by the presence of a high interstellar radiation field (ISRF) and a high cosmic ray ionisation rate (CRIR). Numerical studies of stellar birth typically neglect this fact, and those that do not have thus far been limited in scope. In this work we present the first comprehensive analysis of hydrodynamical simulations of star formation in extreme environments where we have increased the ISRF and CRIR to values typical of the CMZ and starburst galaxies. We note changes in the fragmentation behaviour on both the core and stellar system scale, leading to top-heavy core and stellar system mass functions in high ISRF/CRIR clouds. Clouds fragment less on the core scale, producing fewer but more massive cores. Conversely, the cores fragment more intensely and produce richer clusters of stellar systems. We present a picture where high ISRF/CRIR clouds fragment less on the scale of cores and clumps, but more on the scale of stellar systems. The change in fragmentation behaviour subsequently changes the mass function of the stellar systems that form through enhanced accretion rates.
\end{abstract}


%%%%%%%%%%%%%%%%%%%%%%%%%%%%%%%%%%%%%%%%%%%%%%%%%%

%%%%%%%%%%%%%%%%% BODY OF PAPER %%%%%%%%%%%%%%%%%%

\section{Introduction} \label{sec:intro}

Simulations of star formation have progressed considerably in their ability to reproduce the observed properties of stars and star-forming regions in recent years. They can accurately re-create the observed initial and core mass functions \citep{krumholz_radiation-hydrodynamic_2012, bate_stellar_2012, maschberger_relation_2014, guszejnov_starforge_2021, mathew_role_2023} as well as the properties of nearby molecular clouds as a whole \citep{vazquez-semadeni_molecular_2007, walch_silcc_2015, girichidis_silcc_2016, tres_simulations_2021}. However, almost all studies neglect one key fact: the majority of stars did not form in nearby molecular clouds.
 
Instead, most star formation occurred in environments akin to starbursts earlier in the Universe's evolution, i.e most star formation occurred alongside significantly elevated star formation rates. Less than 25\% of stars formed in the present ($z \rm < 0.7$) universe. Instead, near 50\% of star formation occurred at redshifts $z \rm > 1.3$ \citep{madau_cosmic_2014}. This was concentrated in starburst and ultra-luminous infra-red galaxies (ULIRGs) where star formation rates could exceed $\rm 1,000 \, M_\odot \, yr^{-1}$ \citep{wild_star_2020}. Even in the present-day Milky Way, a significant amount of star formation is occurring in the Central Molecular Zone (CMZ), a vastly different environment to nearby clouds \citep{henshaw_protostars_2023}. There are suggestions that this environment is similar to the starburst galaxies where we believe most star formation to have taken place \citep{kruijssen_comparing_2013}. Such environments are presumed to have elevated cosmic ray ionization rates (CRIRs) and interstellar radiation fields (ISRFs) compared to that of the solar neighbourhood, as a direct result of their enhanced star formation. Intense star formation produces numerous O/B type stars that emit ionising radiation, which later die and produce supernova remnants which are important sources of cosmic rays via diffusive shock acceleration \citep{krymskii_regular_1977, axford_acceleration_1977, bell_acceleration_1978, bell_acceleration_1978-1, blandford_particle_1978, krumholz_cosmic_2023}. Cosmic ray ionisation rates can exceed 100-1,000 times that of the solar neighbourhood in these environments \citep{papadopoulos_cosmic-ray_2010, le_petit_physical_2016, ginsburg_dense_2016}. In addition, observations of starburst galaxies and ULIRGs suggest they experience an enhanced ISRF in addition to an increased CRIR \citep{stacey_158_1991, davies_molecular_2003, shangguan_interstellar_2019} that scales with their rate of star formation \citep{rowlands_evolution_2015}. For the CMZ, simulations of clouds there suggest an ISRF and CRIR 100-1,000 times solar is needed to reproduce their observed properties \citep{clark_temperature_2013}. 

Increasing these parameters has an impact on the thermodynamic state of molecular clouds. Observations of the galactic centre suggest that its dense gas is much warmer than in nearby clouds \citep{ao_thermal_2013, ginsburg_dense_2016}, and simulations corroborate this \citep{wolfire_neutral_2003}. This is important as the thermodynamic state of a molecular cloud has a role in setting the initial mass function (IMF): a warmer cloud has a larger Jeans mass, which sets the peak of the IMF \citep{bate_origin_2005, larson_thermal_2005, bonnell_gravitational_2008, klessen_physical_2016}. It is thought that the thermodynamic behaviour of clouds overall has an impact on star formation \citep{li_effects_2003, jappsen_stellar_2005}, which in turn is modified by an increased ISRF and CRIR. \citet{glover_is_2012} have suggested that star formation itself is dependent on interstellar gas' ability to shield itself from the ISRF, and an increased ISRF is more difficult to shield against.

These arguments suggest that star formation may be different in these extreme environments, and evidence for this already exists. \citet{motte_unexpectedly_2018} found a notably shallow core mass function (CMF) in a region of the galactic bar experiencing a burst in star formation, and both \citet{lu_stellar_2013} and \citet{hosek_unusual_2019} have presented the case for a top-heavy IMF in the galactic centre. While the IMF is generally thought to be universal, these studies provide hints that it may not be and through sampling of the IMF, \citet{dib_massive_2017} suggested that it has an underlying distribution broad enough to allow for it to be non-universal. If the IMF and CMF in these environments significantly deviates from those in the solar neighbourhood, it would mean that a significant proportion of stars formed with an IMF unlike that estimated from star formation in the solar neighbourhood.

Studies that focus on circumstances unlike contemporary star formation in the solar neighbourhood have thus far focused on low-metallicity and Population III star formation \citep{clark_formation_2011, greif_simulations_2011, dopcke_initial_2013, bate_statistical_2014, bate_statistical_2019}. \citet{clark_does_2015} and \citet{clark_tracing_2019} have simulated some variation of the ISRF and CRIR, but not in the context of star formation itself. \citet{klessen_stellar_2007} investigated the stellar mass spectrum in starburst environments, but used a simple polytropic equation-of-state rather than a more physical treatment of the ISM, and did not resolve star formation below $\rm 1 \, M_\odot$. \citet{guszejnov_effects_2022} and \citet{guszejnov_effects_2023} vary the ISRF in the context of star formation, but do not investigate the fragmentation and core mass function of their clouds, nor do they sample the full range of plausible ISRFs. \citet{whitworth_minimum_2024} have explored the effect of the ISRF and CRIR on the low mass end of the IMF, but did not do so with a full hydrodynamic model. 

In this work we investigate how the ISRF and CRIR impact the fragmentation, mass functions and star formation rates of molecular clouds, in the context of understanding how star formation may diverge from that in the solar neighborhood in extreme environments. In Section \ref{sec:methods} we describe the simulation setup and initial conditions, in Section \ref{sec:cloudProperties} we present the general properties and thermodynamic state of the simulated clouds, in Section \ref{sec:sinks} we investigate the properties of the sink particles formed in the clouds, in Section \ref{sec:fragmentation} we study the fragmentation of the clouds and the corresponding core mass function, in Section \ref{sec:crAttenuation} we assess the impact of cosmic ray attenuation, and in Section \ref{sec:discussion} we discuss the implications and limitations of this study.  
 
\section{Methods} \label{sec:methods}

\subsection{Numerical Simulations}

The simulations presented in this work were performed using the adaptive mesh refinement (AMR) code \textsc{arepo} \citep{springel_e_2010}. It uses the Voronoi tessellation of a set of free-to-move points to generate a mesh for which the equations of ideal hydrodynamics are solved using a second-order finite volume scheme. A particular advantage of \textsc{arepo} is that it allows for continuous refinement and de-refinement through the insertion or deletion of mesh-generating points. We use this feature to enforce Jeans refinement, where we ensure that there are always 16 cells per local Jeans length, in order to keep collapsing regions sufficiently resolved. We have made a number of changes to the version of \textsc{arepo} presented in \citet{springel_e_2010}, which are described in detail by \citet{hunter_towards_2023}. For completeness we will summarise them below.

\subsubsection{Sink Particles}

We include sink particles \citep{bate_modelling_1995, federrath_modeling_2010} as a computational aid and as a representation of stellar systems in the simulations. We use the same algorithm for creating sink particles as \citet{hunter_towards_2023}, where sink particles are inserted in regions that are: 1) sufficiently dense, 2) actively collapsing to a central point, 3) in a minimum of gravitational potential, 4) sufficiently isolated from other sinks and 5) gravitationally bound. We insert sink particles above a minimum density of $\rm 1.991 \times 10^{-16} \, g\,cm^{-3}$ and use an interaction radius of $\rm 180 \, AU$. The interaction radius defines the size of the region that must meet the above conditions, and the radius within which the sink can accrete material. This can be interpreted as the effective size of the sink. The sizes of the sinks we use in this work make them poor analogues of individual stars, and instead they better represent stellar systems of one or more stars. The sink particles are gravitationally softened over a radius of $\rm 40 \, AU$.

\subsubsection{Astrochemistry and Thermodynamics}

The investigation presented here is possible thanks to the astrochemistry network \textsc{sgchem} that has been continuously developed for \textsc{arepo}. In brief, it models the chemical evolution of the ISM using a modified version of the chemical network outlined in \citet{gong_simple_2017}; details of the modifications can be found in \citet{hunter_towards_2023}. It accounts for the attenuation of the ISRF by dust shielding and also the self-shielding of $\rm H_2$, C and CO, and includes heating and cooling processes for various atomic and molecular species \citep{clark_tracing_2019}. While the astrochemistry is of secondary importance to this work, the inclusion of ISRF attenuation, cosmic ray ionisation and dust heating and cooling allows us to model high radiation environments (in lieu of a full radiative hydrodynamics model). In two of the simulations, we model a simple column-density dependent cosmic ray attenuation using the treatment outlined in Appendix \ref{sec:crAttenuationModel} alongside the standard physics described here.

\subsection{Initial Conditions} \label{sec:ics}

We initialise the simulations as uniform density spherical clouds of radius $\rm 4.10$ pc. Each cloud contains $\rm 1 \times 10^4 \, M_\odot$ of gas, for an initial number density of $n = \rm 10^3 \, cm^{-3}$. The gas and dust in the cloud are initially at $\rm 40 \, K$ and $\rm 15 \, K$ respectively. These temperatures adjust quickly to their equilibrium values once the simulations begin, thus the specific values are of little consequence. We place the clouds inside a box with side lengths 7.5 times their diameter and populate the remainder of the box with gas with an initial number density of $\rm 10 \, cm^{-3}$ and temperature of $\rm 80 \, K$. We begin the simulations with $\rm 1 \times 10^6$ mesh-generating points, for an initial mass resolution of $\rm 0.01 \, M_\odot$.

Molecular clouds are known to be the sites of turbulent cascades that are inherited from their wider environment \citep{larson_turbulence_1981}. Therefore we apply a turbulent velocity field to the cloud with a power spectrum  $P(k) \propto k^{-4}$ with a natural mix of compressive and solenoidal modes as in \citet{federrath_comparing_2010} and \citet{lomax_effects_2015}. We scale the energy of these turbulent motions so that the simulated cloud is initially in approximate virial equilibrium, $E_{\rm K} = \frac{1}{2} |E_{\rm G}|$, where $E_{\rm K}$ is the total kinetic energy of the cloud and $E_{\rm G}$ the total gravitational potential energy of the cloud. Once the simulations begin the turbulence is left to freely decay in shocks. We use an isolated, spherical, turbulent cloud as this simplifies the initial conditions, and it has been shown to reproduce the properties of observed clouds well \citep{priestley_simulated_2023}. However the density, geometry and velocity dispersion of the initial conditions are more typical of the solar neighbourhood than of clouds in the CMZ or starbursts \citep{imara_alma_2019,dale_dynamical_2019,henshaw_protostars_2023}. However the clouds in the high-$\rm \gamma_{SFR}$ begin to show greater resemblance to those of the CMZ or starbursts as the simulations progress.

We use four combinations of the ISRF and CRIR, each increasing by an order of magnitude in turn. The lowest values of the CRIR and ISRF, $\rm{G = 1.7} \, G_0$ and $\xi_{\rm H_2} \rm = 3.5 \times 10^{-16} \, s^{-1}$ (together adopted as the fiducial case, labelled $\rm \gamma_1$), correspond to estimates of the contemporary values in the solar neighbourhood \citep{draine_photoelectric_1978, indriolo_investigating_2012} and the highest values (labelled $\rm \gamma_{1000}$) are $\rm 10^3$ times higher. From here on we will refer to the combination of the ISRF and CRIR as $\rm \gamma_{SFR}$, to simplify discussion and reinforce the variables as proxies for a changing external star formation rate. We run two simulations for each value of $\rm \gamma_{SFR}$, using two different turbulent seeds. Results in the following sections are typically shown for one turbulent seed only, unless the results are strongly influenced by the seed in which case both runs are shown. We also run two separate simulations where cosmic ray attenuation is modelled, using the same initial conditions as $\rm \gamma_1$ and $\rm \gamma_{1000}$. Results from these runs are typically included in Figures, but are discussed separately from the standard clouds in Section \ref{sec:crAttenuation}. The details of each simulation's initial conditions are shown in Table \ref{tab:simulations}. 

The $\rm \gamma_{SFR}$ value also influences the equilibrium abundance of each of the chemical species modelled in the network, thus we begin each simulation with slightly different abundances. We determine the equilibrium abundances by evolving a low-resolution simulation for each $\rm \gamma_{SFR}$ until chemical abundances stabilise. We set the initial abundance of each species to the corresponding abundance at a density of $\rm 10^3 \, cm^{-3}$ in the test simulations (see Table \ref{tab:initSpecies}). We use the same initial abundances for both turbulent seeds, and for the runs including cosmic ray attenuation. For the total abundances of carbon and oxygen relative to hydrogen, we use the values $x_{\rm C} \rm= 1.4 \times 10 ^{-4}$ and $x_{\rm O} \rm = 3.2 \times 10^{-4}$ as suggested by \citet{sembach_modeling_2000}.  

\begin{table}
    \caption{The simulation labels and their respective parameters. $M_0$ is the total mass of the cloud, $R_{\rm c}$ is the initial spherical radius of the cloud. G and $\xi_{\rm H_2}$ describe the ISRF and CRIR (in \citet{habing_interstellar_1968} and SI units respectively), with values for the fiducial case ($\gamma_1$) set to estimates of the values obtained in the solar neighbourhood \citep{draine_photoelectric_1978, indriolo_investigating_2012}. $x_{\rm H_2}$ is the initial abundance of molecular hydrogen.} 
    \centering
    \begin{tabular}{c|c|c|c|c|c}
        \hline
         Simulation    & $M_0$ [$\rm M_\odot$] & $R_{\rm c}$ [pc]& $G \, [\rm G_0$] & $\xi_{\rm H_2}$ [$\rm s^{-1}$] & $x_{\rm H_2}$\\
        \hline \hline
         $\gamma_1$      & $1\times 10^4$    & 4.10       & 1.7       & $3.5\times 10^{-16}$ & 0.358 \\
         $\gamma_{10}$   & $1\times 10^4$    & 4.10       & 17        & $3.5\times 10^{-15}$ & 0.382 \\
         $\gamma_{100}$  & $1\times 10^4$    & 4.10       & 170       & $3.5\times 10^{-14}$ & 0.254 \\
         $\gamma_{1000}$ & $1\times 10^4$    & 4.10       & 1,700     & $3.5\times 10^{-13}$ & 0.046 \\
         \hline
    \end{tabular}
    \label{tab:simulations}
\end{table}

\section{Cloud Properties} \label{sec:cloudProperties}

\subsection{Cloud Structure and Morphology}

\begin{figure*}
    \centering
    \includegraphics[width=\textwidth]{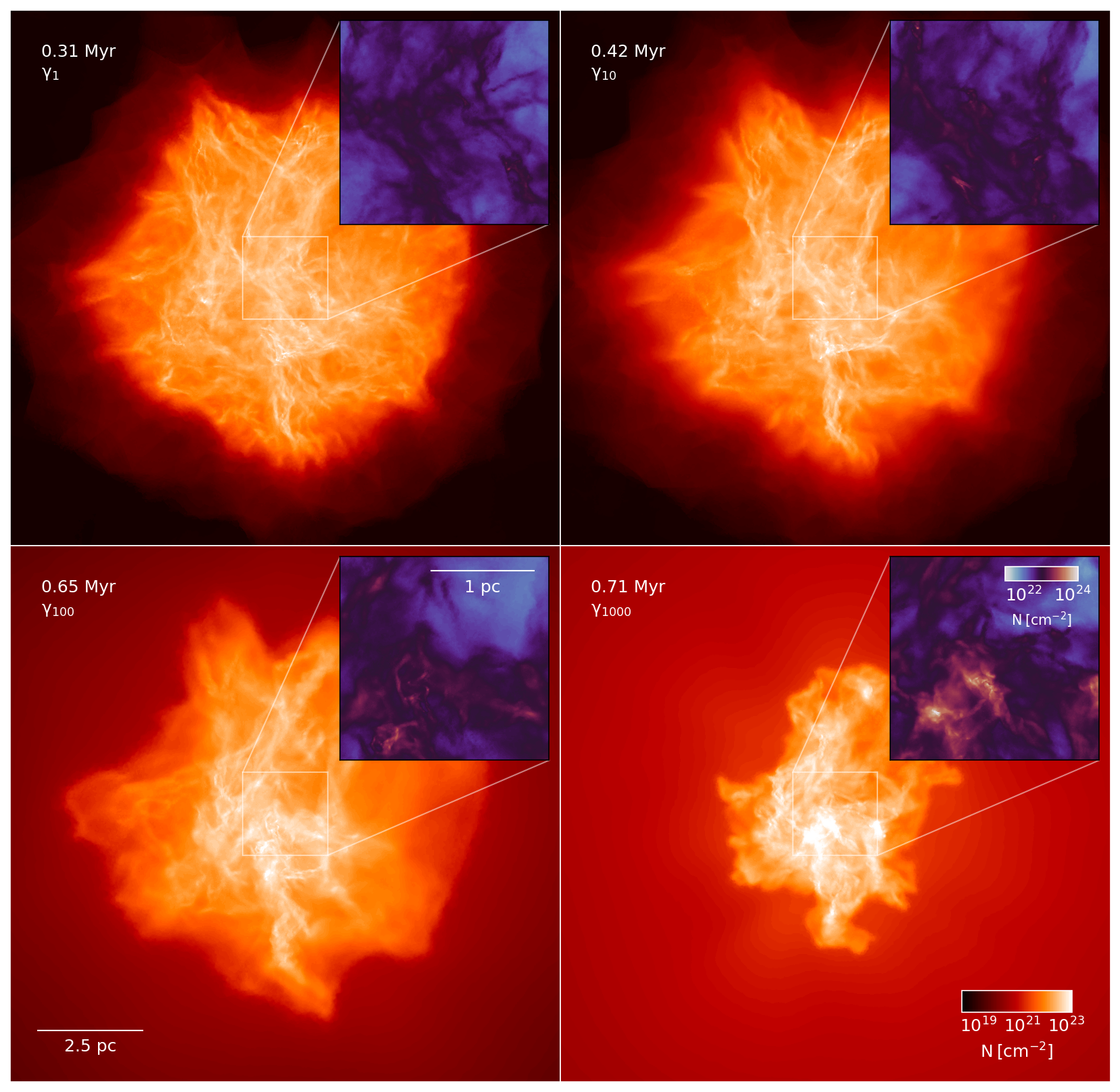}
    \caption{Column density maps of each cloud at the onset of sink formation (the time of which is shown in the top-left corner alongside the simulation identifier). Insets show a 2x2pc region zoomed into the cloud centres.}
    \label{fig:wholeCloudOnsetSFR}
\end{figure*}

The simulations were terminated after $\rm 1.39 \, Myr$ (1 free fall time), by which time each cloud had turned between 2,000 and 4,000 $M_\odot$ of gas into sinks. Figure \ref{fig:wholeCloudOnsetSFR} shows column density maps of the clouds at the onset of sink formation along with the corresponding time. Increasing $\rm \gamma_{SFR}$ changes both the internal and external structure of the clouds as well as the time of the onset of sink formation.

Externally, the transition region between the cloud and its surroundings increases in extent and density as more outlying gas either expands or escapes the cloud's potential well. This "boiling off" of the cloud's outer envelope is a result of the additional heating from the ISRF; the extra energy from this heating causes the gas to either expand and cool (increasing the extent of the transition region) or to escape the cloud entirely (creating a denser surrounding medium). In the case of $\gamma_{1000}$ this effect is so strong that the outer gas begins to exert a pressure back onto the cloud, compressing it. A similar result was noted by \citet{clark_does_2015} when using an enhanced ISRF and CRIR. The contrast between the outlying envelope and the cloud also becomes stronger for the high-$\rm \gamma_{SFR}$ clouds, as the low density gas expands uniformly into the surrounding region and no longer clings to the edges of the clouds as it does in the low-$\rm \gamma_{SFR}$ clouds. As a result of this the division between cloud and envelope is much clearer in the high-$\rm \gamma_{SFR}$ simulations.

Internally, the structures that form within the clouds become larger as $\rm \gamma_{SFR}$ increases. The low-$\rm \gamma_{SFR}$ clouds contain complex, rich, networks of thin filaments and cores, whilst by contrast the high-$\rm \gamma_{SFR}$ clouds show larger, denser clumps. This too is a result of additional heating from the ISRF and the resulting increase in the sound speed of the gas. This weakens the ability of turbulent shocks to produce density contrasts, and increases the local Jeans mass. These effects compound and effectively stabilize the cloud, preventing the formation of small-scale structure and requiring structures to become more massive before they can collapse. This has the additional effect of delaying sink formation, as more time is needed to assemble the massive structures that can trigger self-collapse.

\subsection{Thermodynamics of the Cloud}

\begin{figure}
    \centering
    \includegraphics[width=\columnwidth]{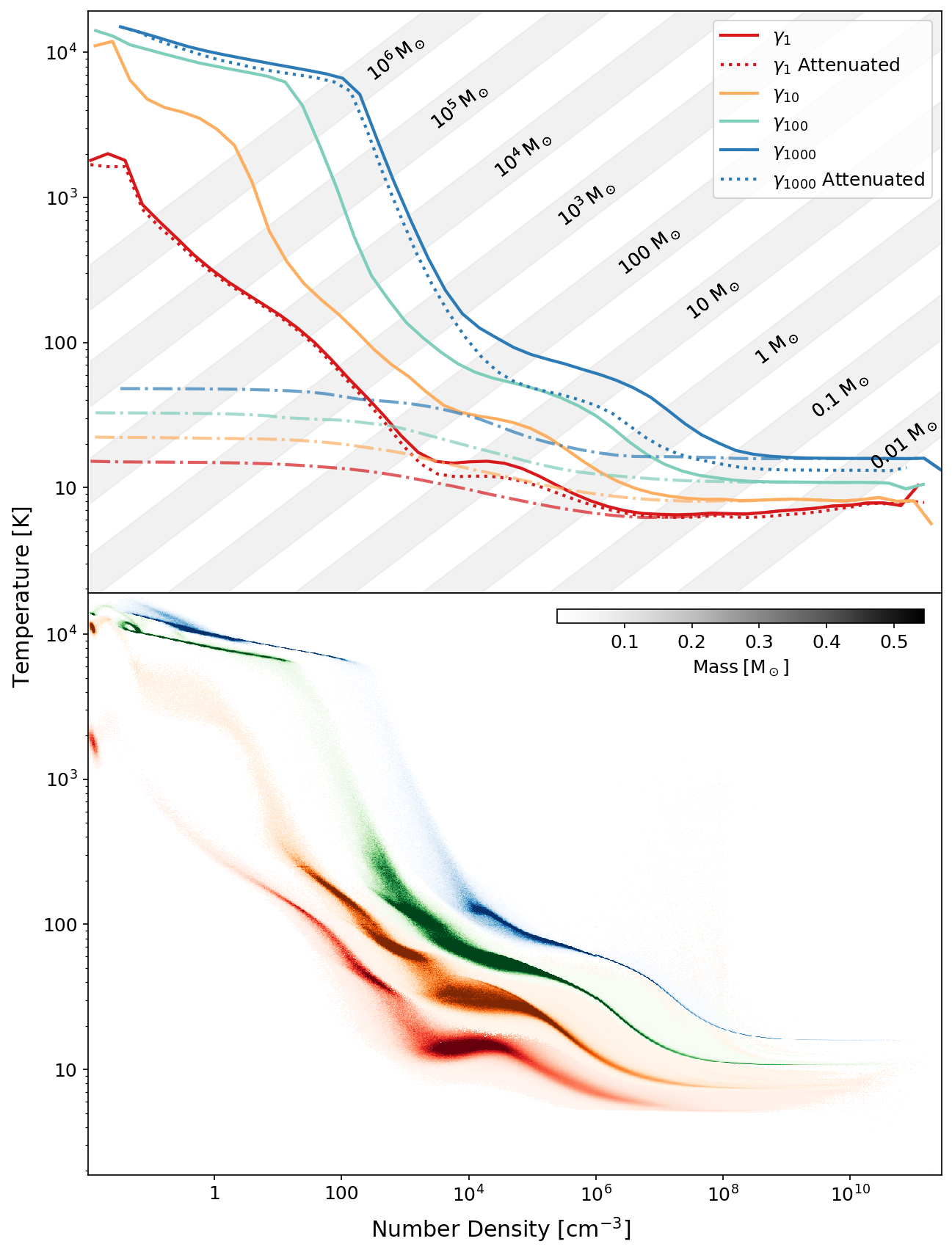}
    \caption{Temperature-density diagrams of the clouds when the simulations were terminated. The top panel shows mass-weighted average gas (solid lines) and dust (dot-dashed lines) temperatures as a function of number density. Grey bands show lines of constant Jeans mass where the spread corresponds to the range of typical mean molecular weights in the ISM ($\mu = 1.4 - 2.4$). The mean molecular weight is likely to vary in a systematic way with varying temperature and density, as the gas becomes fully molecular at high densities (increasing $\mu$). Gas temperatures for the runs with attenuated cosmic rays are indicated with dotted lines. The bottom panel shows the raw distribution of cell gas temperatures against number density, weighted by the mass of each cell.}
    \label{fig:tempDensityEnd}
\end{figure}

In Figure \ref{fig:tempDensityEnd} we show temperature-density diagrams of the clouds in both binned and raw forms for the gas (both panels) and the dust (top panel only). Number density in this Figure and all subsequent Figures is the number density of hydrogen nucleons in all forms. Increasing $\rm \gamma_{SFR}$ increases the temperature of the gas and dust across all densities. The gas and dust temperatures increase monotonically with the increase in $\rm \gamma_{SFR}$, except at very low densities. 

Increasing $\rm \gamma_{SFR}$ also pushes the transition out of the warm neutral medium (WNM) to higher densities. While $\gamma_1$ does not reach temperatures of $\rm 10^4 \, K$ at all during the simulation, $\gamma_{1000}$ does so for densities up to $\rm 10^3 \, cm^{-3}$. At such a density we usually expect the gas to have become cold and molecular due to attenuation of the ISRF. Instead, the ISRF is so intense that shielding is not strong enough to allow the gas to cool yet, as it is in the lower-$\rm \gamma_{SFR}$ clouds. Therefore the clouds remain hot until a (higher) density where shielding can attenuate the strengthened ISRF. \citet{wolfire_neutral_2003} similarly found that the WNM-CNM transition is pushed to higher densities when modelling the Galactic Centre, due to an increased ISRF/CRIR. The result of this cooling delay is that more hot gas "boils off" the cloud, as we saw in Figure \ref{fig:wholeCloudOnsetSFR}, producing a dense halo that can push back onto the cloud. It should be noted, however, that the behaviour of the gas at densities below the starting density ($\rm 10^3 \, cm^{-3}$) will be impacted by the choice of initial conditions and may not be representative of the typical ISM. 

Another effect of increasing $\rm \gamma_{SFR}$ is that the transition from the cold neutral medium (CNM) to molecular gas is disrupted. This is a direct consequence of change to the WNM, as the high-$\rm \gamma_{SFR}$ clouds are still hot at the typical densities of the CNM. Instead, these clouds transition straight to molecular gas as they cool and have no stable cold atomic regime (see Figure \ref{fig:fracAbunds} for the abundances of hydrogen and carbon species against density). An implication of this is that the density where both $\rm H_2$ and CO form both increases and converges with $\rm \gamma_{SFR}$. The two begin to form in tandem in the high-$\rm \gamma_{SFR}$ clouds, reducing the amount of CO-dark molecular gas we see in the clouds. This may make CO a better tracer of molecular gas in high-SFR environments.

The transition from the WNM to the CNM and then to molecular gas occurs as a result of shielding from the ISRF increasing as the optical depth of a cloud increases. The amount of shielding needed to transition from the CNM to molecular gas is slight, but in a low-density cloud the CNM and molecular regime can coexist thanks to the gradual change in shielding with cloud depth at low densities. However, in the high-$\rm \gamma_{SFR}$ clouds the density at which the clouds leave the WNM is much higher. Shielding now increases rapidly with cloud depth, so that even shallow depths into the cloud exceed the shielding needed to produce molecular gas. Therefore no CNM is able to form. 

Increasing $\rm \gamma_{SFR}$ increases the density at which dust-gas coupling occurs. For the gas and dust to thermally coupled, the energy transfer from gas-gain interactions must be large enough to regulate the gas temperature. This energy transfer increases with the square of number density, so dust-gas coupling typically occurs at high densities. When we increase $\rm \gamma_{SFR}$, we increase the heating rate of the gas, but only slightly increase the efficiency of gas-grain collisions (as it has a dependence on temperature. The dust no longer regulates the gas temperature until a higher density where the energy transfer is large enough to counteract the additional heating. This results in a delay of nearly two orders of magnitude in the coupling density between $\gamma_1$ and $\gamma_{1000}$, resulting in the high-$\rm \gamma_{SFR}$ clouds remaining at a higher temperature at higher densities and not achieving an isothermal state until later. This has implications for the fragmentation of the clouds; higher temperatures increase the Jeans mass which sets the scale over which fragmentation occurs \citep{jappsen_stellar_2005, palau_gravity_2015}. Fragmentation only slows when the clouds become isothermal and the point at which that happens is impacted by $\rm \gamma_{SFR}$.

The temperature of the dust, while stable compared to that of the gas, is increased by $\rm \gamma_{SFR}$. Dust temperatures increase in proportion to $\rm \gamma_{SFR}$, but differ by only $\rm 15 \, K$ across all clouds at the sink creation density. This still corresponds to an order of magnitude difference in the Jeans mass, however, and will affect the formation of the sink particles.

% RE: REVIEWERS COMMENTS

Another consequence of increased dust temperatures is that the far-infrared spectral energy distribution (SED) of the clouds will be different. Starburst galaxies observed at high redshifts naturally have a SED that is shifted to shorter wavelengths, thanks to the warmer cosmic microwave background at earlier epochs. Figure \ref{fig:tempDensityEnd} shows that dust temperatures are hotter in these environments too, blue-shifting the SED further and changing the observability of such systems \citep{de_rossi_far-infrared_2018}). This provides an observational signature for high-$\rm \gamma_{SFR}$ environments that can be cross-examined with known starburst systems and their properties.

\subsection{Dominant Thermal Processes}

\begin{figure*}
    \centering
    \includegraphics[width=\textwidth]{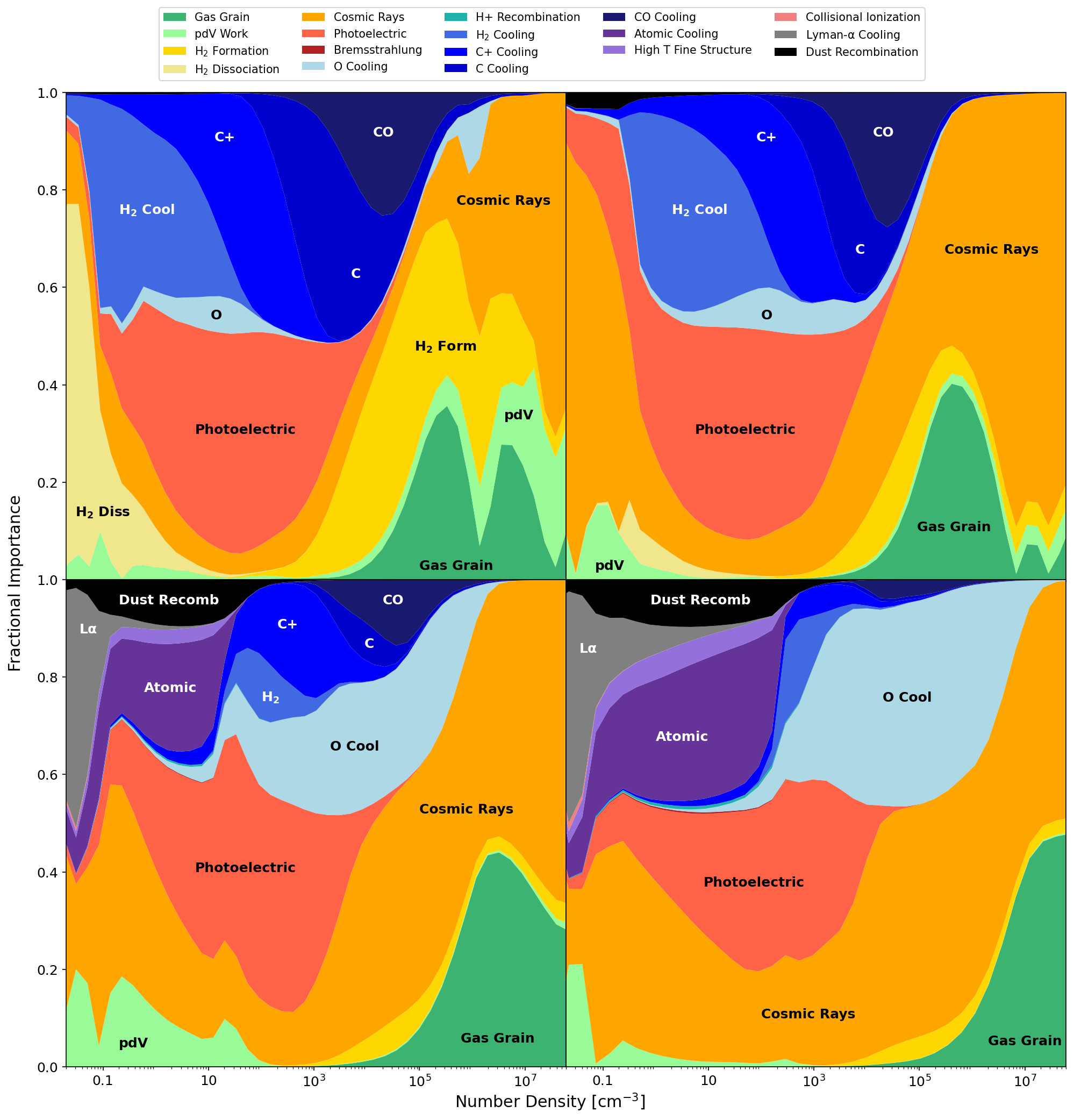}
    \caption{The relative heating and cooling rates included in the chemical network and their fractional importance against number density. Rates are binned into equal width bins of number density and a mass-weighted average rate is found for each bin/rate. The rates are then normalised to account for the changing sum of raw rate values with number density. Rates are coloured depending on whether they heat (reds and oranges), cool (blues and purples) or can do either (greens). The panels are oriented as previous figures.}
    \label{fig:heatCoolFractional}
\end{figure*}

The explanation for the differing temperature-density structures of the clouds is the changing importance of different heating and cooling rates. Increasing $\rm \gamma_{SFR}$ strengthens certain heating processes, while also affecting the abundances of the chemical species in the network we are using. Understanding how these changes interact reveals why the thermodynamic state of the clouds differs so strongly. Therefore, we show the fractional importance of the different heating and cooling rates against number density in Figure \ref{fig:heatCoolFractional}. 

Photoelectric heating is a dominant process in all the clouds, but for high-$\rm \gamma_{SFR}$ clouds there is an increasingly large contribution from cosmic ray heating. The source of these (the ISRF and CRIR respectively) both increase in step as we increase $\rm \gamma_{SFR}$, so we might expect these heating sources to also increase in step. However, each of these has a dependence on the chemical state of the gas, and the physical structure of the clouds. The efficiencies of both cosmic ray and photoelectric heating are modulated by the abundance of atomic hydrogen, and photoelectric heating becomes less effective as clump and core sizes increase (thus increasing attenuation for the same density). These complex dependencies interact to allow cosmic ray heating to displace photoelectric heating as the most important heat source in the high-$\rm \gamma_{SFR}$ clouds. At the highest densities photoelectric heating becomes completely irrelevant, as the ISRF is fully attenuated no matter the core or structure size. Cosmic rays are the primary heat source, balanced by gas-grain cooling once the gas and dust are thermally coupled.

Cosmic rays become progressively more important at low densities, yet appear anomalously so for $\rm \gamma_{10}$. This is not because the cosmic rays are stronger than in $\rm \gamma_{100}$, but rather that they are not yet being balanced by any cooling process. The cosmic ray heating is not yet intense enough to heat the gas to temperatures that facilitate Lyman-$\alpha$ and other WNM cooling processes\footnote{Including atomic cooling, which in the network refers to the emission resulting from the collisional excitation of various atomic and ionic species, assuming collisional ionisation equilibrium.}. $\rm H_2$ cooling is also inactive here as the enhanced cosmic ray rate results in the complete dissociation of molecular hydrogen at lower densities. This causes $\rm \gamma_{10}$ to seem as if it is being heated more intensely, even though it is not. $\rm \gamma_{100}$ and above are experiencing stronger heating, such that the gas has become hot enough to activate high-temperature cooling processes to balance it.

Increasing $\rm \gamma_{SFR}$ changes the dominant coolant at intermediate densities. Where in the fiducial cloud cooling is dominated by various carbon-containing species (C$^{+}$, C, CO), in runs $\rm \gamma_{100}$ and $\rm \gamma_{1000}$ atomic oxygen is the most important coolant. The transition between different carbon species also becomes more rapid and shifted to higher densities in the higher $\rm \gamma_{SFR}$ clouds. The different species of carbon are useful for tracing clouds at different evolutionary stages \citep{clark_tracing_2019} however, it appears that this is no longer possible for high-$\rm \gamma_{SFR}$ clouds. Oxygen becomes a more important coolant thanks to the inability of CO to form: increased temperatures and cosmic-ray dissociation act to reduce the abundance of CO \citep{bisbas_effective_2015}, leaving oxygen fine structure cooling to dominate.

We caution that the importance of $\rm H_2$ cooling and dissociation heating at low densities may be an artifact of the initial conditions, as we begin the simulations with some fraction of the gas in molecular form. As the cloud expands, its initial $\rm H_2$ content is retained even in gas orders of magnitude below the density where molecular hydrogen can form. In the low-$\rm \gamma_{SFR}$ clouds the CRIR is not high enough to dissociate this $\rm H_2$ on a short enough timescale and some persists to extremely low densities. This is why there is a strong contribution from $\rm H_2$ cooling and dissociation in $\rm \gamma_1$ and $\rm \gamma_{10}$. These effects do not change the dynamics or fragmentation behaviour of the cloud, only the lowest-density thermodynamics that are of secondary importance to this study.

A final important feature is the ramp-up of gas-grain cooling to balance cosmic ray and $\rm H_2$ formation heating at high densities. This process happens later in the high-$\rm \gamma_{SFR}$ clouds, in accordance with the delayed dust-gas coupling we have mentioned. By the sink creation density, gas-grain cooling is almost perfectly balancing the corresponding heating. This explains the isothermality of this gas at very high densities, the balance between heating and cooling is almost exactly 50/50. The processes we have mentioned also dominate over pdV-work in the higher-$\rm \gamma_{SFR}$ clouds, showing that the thermal and chemical state of the gas is only weakly coupled to its dynamics.

\section{Sink Particles} \label{sec:sinks}

\subsection{Sink Mass Function}

The sink particles used in these simulations have sizes equivalent to or larger than that of typical binary systems \citep{duchene_stellar_2013} and we treat them as stellar systems rather than individual stars. The system mass function (SMF) typically informs and follows the initial mass function \citep{offner_protostars_2014} and the distribution of sink properties reflects how the cloud is fragmenting on stellar scales.

\begin{figure*}
    \centering
    \includegraphics[width=\textwidth]{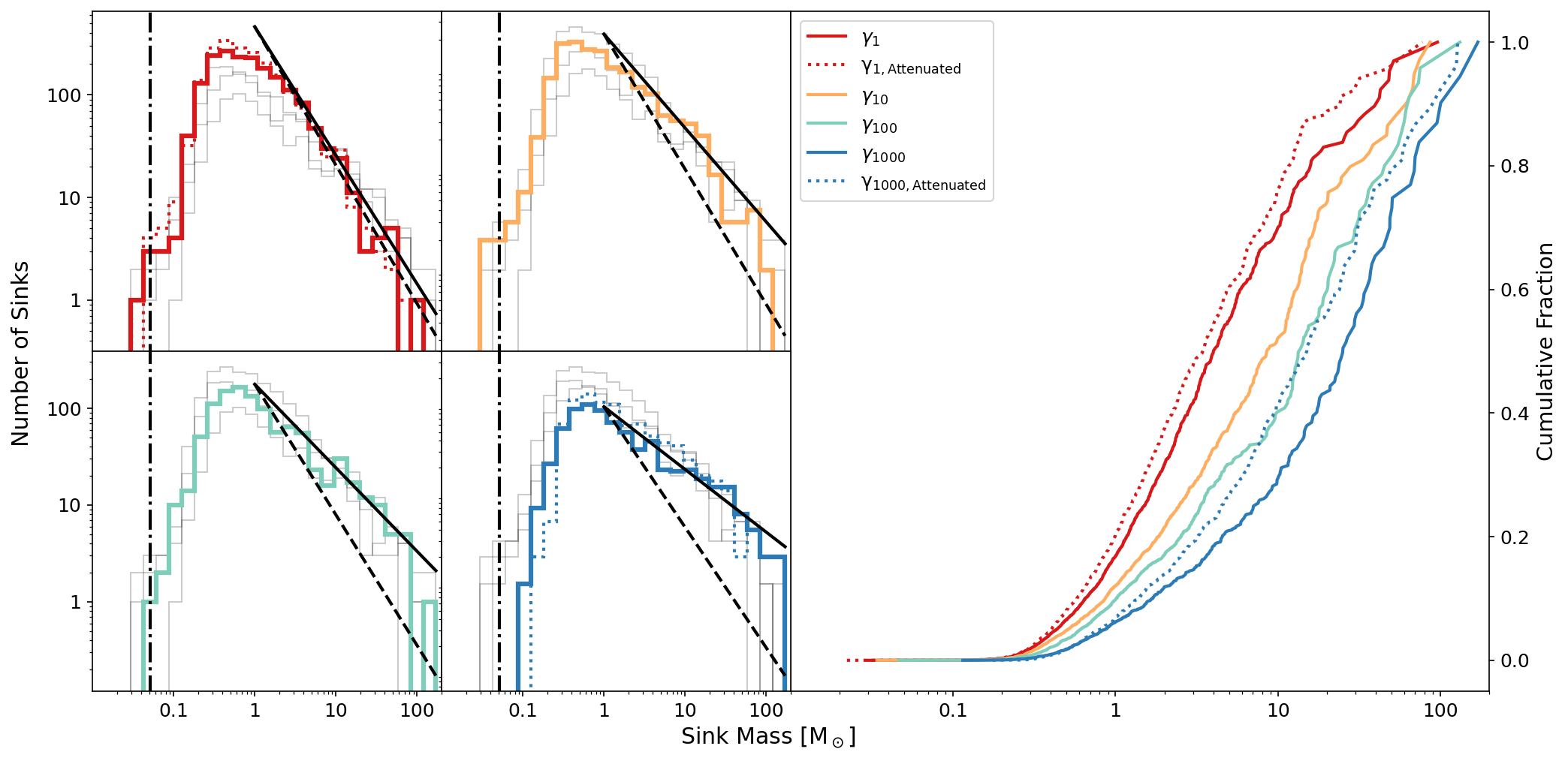}
    \caption{The sink particle mass function of each simulation in histogram and cumulative form. The distributions are sampled when all simulations have reached 3,000 $\rm M_\odot$ in total sink mass. The dashed lines in the left panels are the Salpeter power-law with exponent $\rm \alpha=-1.35$ \citep{salpeter_luminosity_1955} and the solid lines are best-fit power-laws to the high mass ($\rm > 1 \, M_\odot$) tail for the non-attenuated runs. The dot-dashed lines represent the effective resolution limit; the Jeans mass at the sink creation density. Each histogram shows the mass function of the corresponding simulation in colour, with the mass functions of the other simulations shown in light-grey. Dotted lines represent the mass functions of the runs with cosmic ray attenuation.}
    \label{fig:sinkMassFunction}
\end{figure*}

\begin{table*}
    \centering
    \caption{Summary statistics of the sink particle mass function of each simulation. Statistics were calculated when each simulation had formed $\rm 3000 \, M_\odot$ in sinks, as with the Figures. The exponent $\alpha_{\rm Fit}$ corresponds to a least squares fit to the power-law $N(m) \propto m^\alpha$ for masses $M \rm > 1 \, M_\odot$. The quoted uncertainty is the estimated standard error to the fit.}
    \begin{tabular}{c|c|c|c|c|c}
        \hline
        Simulation & $\alpha_{\rm Fit}$ & $M_{\rm Mean} \, \rm [M_\odot]$ & $M_{\rm Med} \rm\, [M_\odot]$ & $M \rm< 1 \, M_\odot$ & $M \rm> 10 \, M_\odot$ \\ \hline\hline
        $\gamma_1$                      & $-1.25 \pm 0.08$ & 1.79 & 0.73 & $16.85\%$ & $30.24\%$  \\ 
        $\rm \gamma_{1, Attenuated}$    & $-1.25 \pm 0.08$ & 1.55 & 0.68 & $20.37\%$ & $23.86\%$ \\
        $\gamma_{10}$                   & $-0.94 \pm 0.08$ & 2.42 & 0.74 & $12.03\%$ & $47.93\%$ \\
        $\gamma_{100}$                  & $-0.86 \pm 0.09$ & 3.05 & 0.78 & $9.79\%$ & $59.83\%$ \\
        $\gamma_{1000}$                 & $-0.70 \pm 0.04$ & 4.69 & 0.92 & $6.20\%$ & $70.85\%$ \\
        $\rm \gamma_{1000, Attenuated}$ & $-0.83 \pm 0.05$ & 3.98 & 1.06 & $6.94\%$ & $58.77\%$ \\
        \hline
    \end{tabular}
    \label{tab:imfStats}
\end{table*}

We show the sink particle mass functions in Figure \ref{fig:sinkMassFunction}, along with best-fit power laws to the high mass tail ($\rm > 1 \, M_\odot$, the properties of which are shown in Table \ref{tab:imfStats}). Increasing $\rm \gamma_{SFR}$ leads to a shallower, top-heavy mass function that deviates from Salpeter. The fiducial cloud, $\gamma_1$, shows good agreement with Salpeter, indicating that the simulation setup reproduces the observed SMF in solar neighbourhood conditions and that changes in the power law are not due to deficiencies in the physics we include. Instead, it is clear that a high $\rm \gamma_{SFR}$ leads to the formation of larger sink particles. The regime where sinks form and accrete is approximately isothermal, and isothermal simulations indicate a dependence of the mass function on the average Jeans mass of a cloud \citep{larson_evolution_1973, bate_origin_2005, bate_dependence_2005, jappsen_stellar_2005}. The gas is hotter and denser in the high-$\rm \gamma_{SFR}$ clouds, and therefore the Jeans mass is larger, leading to larger sink particles.

From the cumulative distributions we can conclude that changes in the SMF are not due to the clouds evolving on different timescales. The maximum sink mass attained by each cloud is similar, yet the distributions are very different. Therefore there has been sufficient time for the sinks to grow to large masses in all clouds, and the distributions are not affected in changes to the evolutionary history of the clouds (e.g $\rm \gamma_1$ beginning sink formation much earlier than $\rm \gamma_{1000}$). Instead, the change in the distributions must be a result of the changing thermal physics of the clouds. 

The peaks of the SMFs are shifted to slightly higher masses as we increase $\rm \gamma_{SFR}$. In addition to this, there is a deficit of low-mass ($\rm < 0.2 \, M_\odot$) sinks in the high-$\rm \gamma_{SFR}$ clouds. Therefore these clouds are bottom-light, and the deficit of low to intermediate mass sinks is further made clear in the cumulative distributions and SMF summary statistics. It has been suggested \citep{larson_thermal_2005,bate_origin_2005} that the peak of the IMF (and by extension the SMF) is affected by thermal physics, and that increasing the temperature increases the peak mass. However, the very slight change in the peak contrasting with the very large change in the slope of the SMF casts some doubt on this idea.

\subsection{Sink Accretion Rates}

\begin{figure}
    \centering
    \includegraphics[width=\columnwidth]{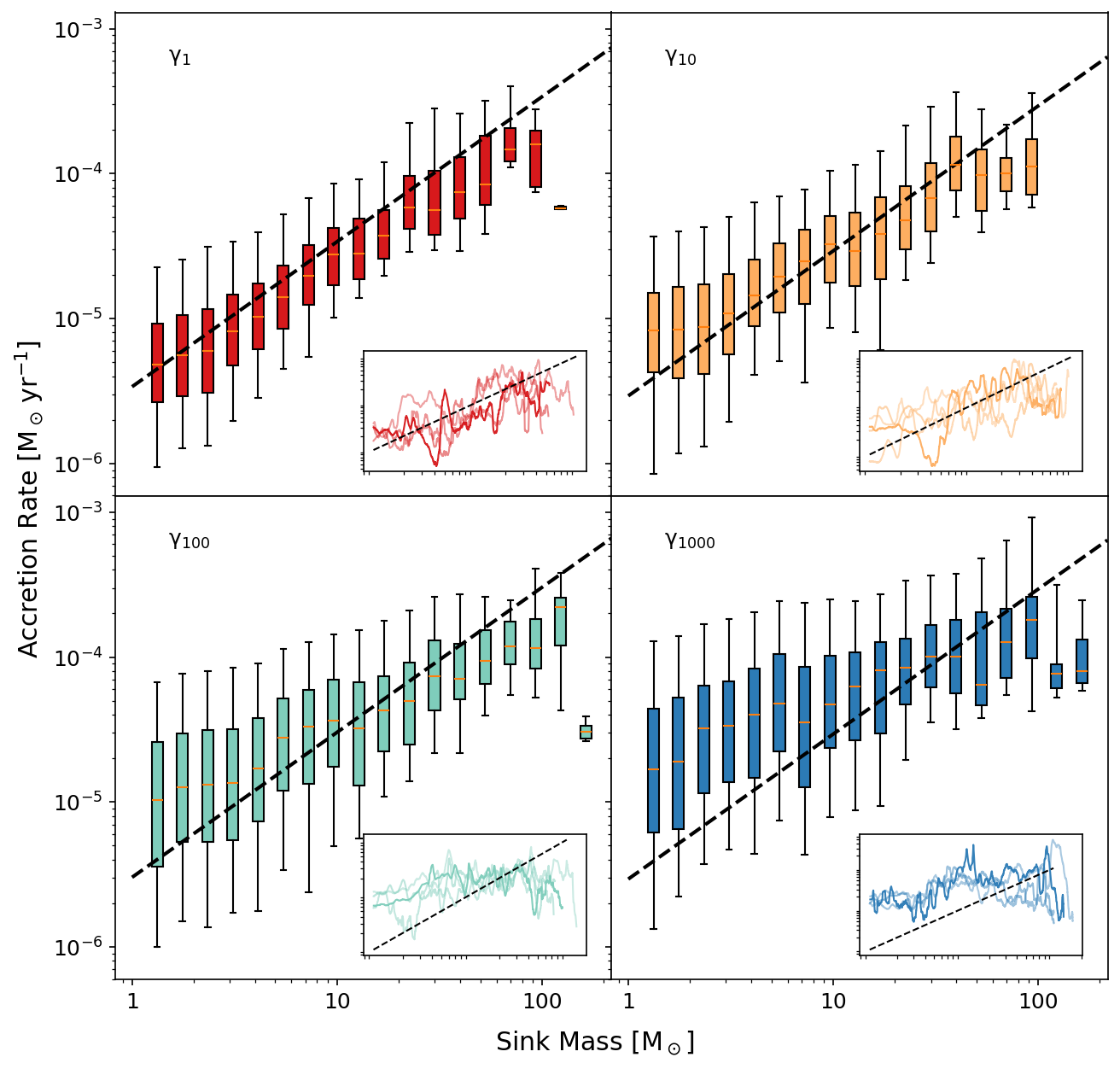}
    \caption{Sink particle accretion rates against sink mass for each simulation for sinks with masses $\rm > 1 \, M_\odot$. Accretion rates are binned by sink mass and a box plot shows the distribution of accretion rates in each mass bin. Accretion rates are sampled for each sink at each snapshot and are calculated from the time it takes for a sink to increase its mass by $\rm 10\%$. The dashed line shows the expected $\dot{m} \propto m$ relationship from \citet{maschberger_relation_2014} and \citet{clark_emergent_2021}. In the bottom-right of each panel we show the individual mass-accretion relationships for the 5 most massive sinks in each simulation. While the box plots contain accretion rates for all sinks at all times, these panels show the evolution of the most massive sinks explicitly.}
    \label{fig:accretionRates}
\end{figure}

The differing system mass functions imply a difference in the growth of the sink particles in each cloud. The sink particles in the high-$\rm \gamma_{SFR}$ clouds must be growing more rapidly than those in the low-$\rm \gamma_{SFR}$ clouds, in order to achieve larger masses in the time between the onset of sink formation and the termination of the runs (which is shorter for the high-$\rm \gamma_{SFR}$ clouds). Figure \ref{fig:accretionRates} shows this to be the case. Accretion rates are elevated for the high-$\rm \gamma_{SFR}$ clouds, particularly for sinks below $\rm 10 \, M_\odot$. At high sink masses, however, they are generally comparable, or less than, their counterparts in the low-$\rm \gamma_{SFR}$ clouds. This suggests a weakening of competitive accretion in the high-$\rm \gamma_{SFR}$ clouds. 

\citet{maschberger_relation_2014} and \citet{clark_emergent_2021} have suggested that an accretion rate proportional to sink mass is needed to reproduce the Salpeter power law. The results presented here support this conclusion, as $\rm \gamma_1$ shows this proportional relationship and is the only cloud to reproduce the Salpeter high-mass tail. When $\rm \gamma_{SFR}$ increases, the relationship between accretion rate and mass shallows, particularly at high sink masses, and a Salpeter mass function no longer emerges. In a highly competitive environment, sinks with greater mass accrete more rapidly as they settle to the dense region at the bottom of the potential well. This leads to a steep accretion-mass relationship where massive sinks grow in proportion to their mass. In the high-$\rm \gamma_{SFR}$ clouds we see a shallower slope, indicating that accretion is not so strongly tied to mass. This implies that accretion may have become less competitive in these clouds, as larger sinks are growing at a similar rate to less massive ones. This inevitably has had an effect on the sink mass function. The inset panels in Figure \ref{fig:accretionRates} show that this is applicable to both individual sinks and sinks in aggregate. The largest sinks in $\rm \gamma_{1}$ accrete at a rate roughly proportional to their mass, whereas in $\rm \gamma_{1000}$ this relationship is approximately flat. Competitive accretion is applicable to the sinks in $\gamma_{1}$ and $\gamma_{10}$, yet breaks down as $\rm \gamma_{SFR}$ increases.

The elevated accretion rates explain why the sink mass function is top-heavy in the high-$\rm \gamma_{SFR}$ clouds; sink particles gain mass more rapidly and therefore end up larger. The flattening of the accretion rate relationship also explains the shallower power law tail. \citet{clark_emergent_2021} suggests that more competitive accretion produces few, massive, systems that then dominate the rest, producing a steep mass function. We would then expect that less competitive accretion would do the reverse, producing a shallower mass function.

\subsection{Sink Formation History}

\begin{figure}
    \centering
    \includegraphics[width=\columnwidth]{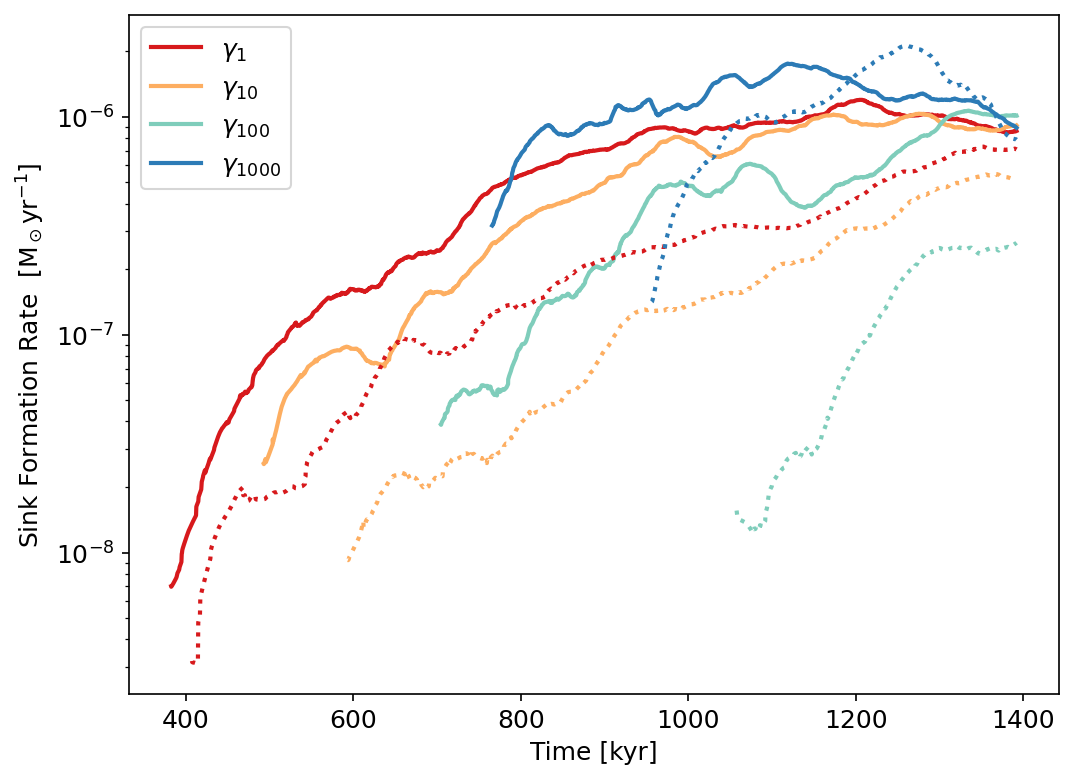}
    \caption{The sink formation rate through time for each of the simulations. The formation rate is calculated as a rolling mean of the change in total sink mass, ${\rm d}M/{\rm d}t$, using a window of 50,000 years. The properties of sink particles are reported every 100 years, so each window covers 500 data points. Solid lines represent the simulations with the first turbulent seed, dotted lines represent the simulations using the second turbulent seed.}
    \label{fig:sinkFormationHistory}
\end{figure}

Figure \ref{fig:sinkFormationHistory} shows the sink formation rate through time for each of the simulations, from the onset of sink formation to the termination of the runs. The sink formation rate is calculated as the mass that either forms new sinks or is accreted by existing sinks. We include both runs in this Figure, as much of the sink formation rate will be strongly influenced by the randomness of turbulent motions \citep{larson_turbulence_1981, klessen_quiescent_2005, ballesteros-paredes_protostars_2007, ballesteros-paredes_gravity_2011}.

Sink formation is delayed as $\rm \gamma_{SFR}$ increases in both simulation runs. This effect is present in both runs, but is more prominent in the secondary run. However with the second turbulent seed $\gamma_{1000}$ begins sink formation earlier than $\gamma_{100}$, whereas with the first turbulent seed $\rm \gamma_{1000}$ begins sink formation very soon after $\rm \gamma_{100}$. This may be due to the crushing by the cloud's outer envelope back onto the cloud, triggering sink formation earlier by artificially speeding up the collapse of the clouds. It is clear that random variation still plays an important role here though, as the $\rm \gamma_{1000}$ clouds do not both begin sink formation before $\rm \gamma_{100}$. 

While the rate of sink formation is subject to randomness, increasing $\rm \gamma_{SFR}$ systematically decreases the sink formation rate significantly in both runs. This is more evident in the first run, where sink formation rates are higher and begin earlier than in the secondary run, but it is a clear effect in both runs. The simulations are all initialised with the same balance between kinetic and gravitational energy, thus the turbulent seed must be responsible for the reduction of sink formation rates in the secondary run. Despite the large differences in sink formation rates between the seeds, the systematic dependence on $\rm \gamma_{SFR}$ persists.

The sink formation rates generally converge to a similar value by the end of the simulation. This suggests that at earlier stages of the clouds' evolution the increased stability of the gas in high-$\rm \gamma_{SFR}$ clouds hampers sink formation. However as the cloud continues to collapse, accretion onto existing sinks may become the limiting factor for the sink formation rate instead.

\section{Fragmentation Properties} \label{sec:fragmentation}

\subsection{Sink Particle Clustering}

\begin{figure}
    \centering
    \includegraphics[width=\columnwidth]{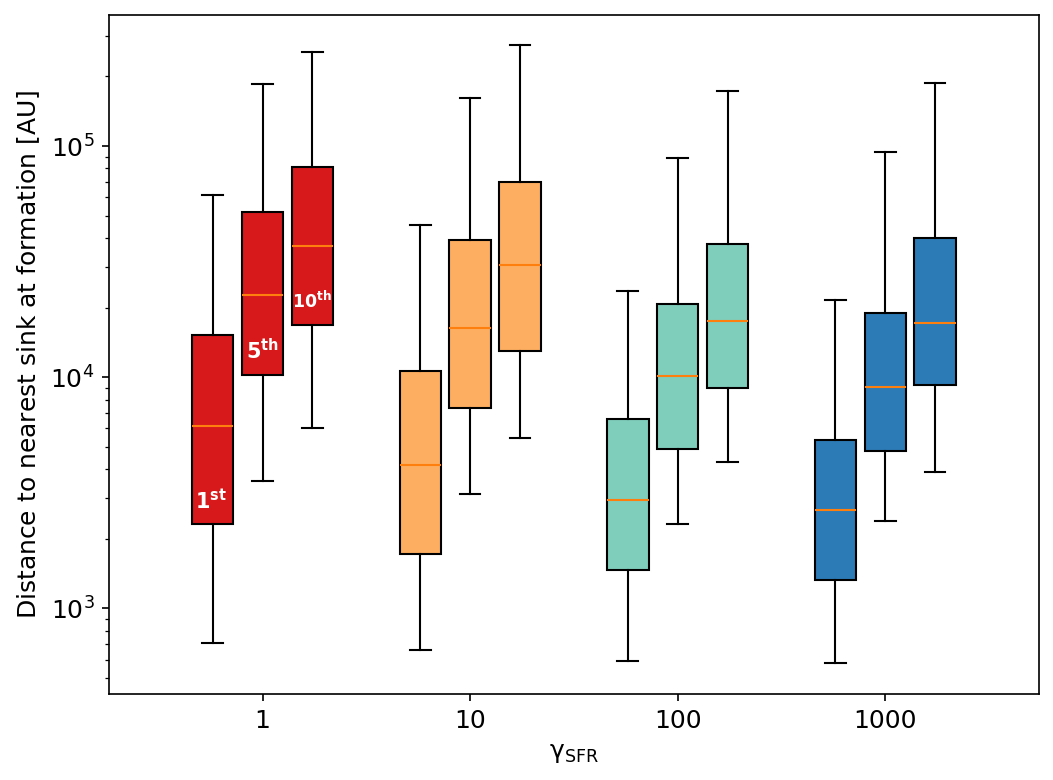}
    \caption{Distributions of the distances to a sinks 1st, 5th and 10th nearest neighbour at the time of its formation. There is thus no data for the first 10 sinks that formed. Outliers are omitted, and the whiskers show the 5th to 95th percentiles of the data.}
    \label{fig:sinkNearestFormation}
\end{figure}

The way that the sink particles cluster, i.e the groups they are formed in, is influenced by the fragmentation of the parent cloud, and the way they grow is impacted by their environment \citep{larson_star_1995,bonnell_massive_2004,bonnell_efficiency_2011,girichidis_importance_2011}. In Figure \ref{fig:sinkNearestFormation} we show the distances to each sink's nearest, 5th and 10th nearest neighbours at the time of its formation. The distance to a sink's nearest neighbour reflects the density of its immediate surroundings when it formed. The closer its $\rm n^{th}$ neighbour is, the more likely it is that the two sinks formed from the same core.

In clouds with a high $\rm \gamma_{SFR}$, the nearest neighbour distances tend to be smaller than low $\rm \gamma_{SFR}$ clouds. This indicates that the sinks in these clouds form in larger groups and a more heavily clustered environment. A typical core ($n = \rm 10^5 \, cm^{-3}$, $T = \rm 15 \, K$) has a Jeans length of around $\rm 10^4 \, AU$, and the median 5th nearest neighbour distance for $\gamma_1$ is around this value, indicating that cores may fragment into 5 or fewer sinks as the nearest neighbour is within the radius of such a core. This is to be expected for the fiducial case, and similar results have been derived through various means \citep{holman_mapping_2013, lomax_effects_2015, ambrose_formation_2024}. The median 10th nearest neighbour distance for $\gamma_{1000}$ is less than $\rm 10^4 \, AU$, indicating that cores may fragment into 10 or more sinks in that cloud. In fact, cores in $\gamma_{1000}$ have a larger Jeans length due to the increased temperature, so cores may contain more than 10 sinks. Thus, increasing $\rm \gamma_{SFR}$ causes cores to fragment into more sinks and produce larger embedded clusters. 

These results indicate that $\rm \gamma_{SFR}$ has changed the fragmentation behaviour of cores, increasing the number of fragments that are produced as they collapse. This suggests that accretion inside the cores should become more competitive and reduce masses \citep{peters_limiting_2010, girichidis_importance_2012}, but we have already seen this to not be the case. The increased fragmentation but lack of competitive accretion would be consistent with each other if the progenitor cores were more massive. Massive cores would then provide each sink with ample gas to accrete before having to compete with those in its surroundings.

\subsection{Core Mass Function}

Identifying the progenitor cores of sinks, and the fragmentation properties of the cloud as a whole, is possible through various means. In the following section we identify cores through a dendrogram analysis. Cores identified with dendrograms allow for a mass to be simply determined and for a core mass function to be constructed. Dendrograms can be derived from either images or 3D simulation data, and are thus common in both numerical and observational studies \citep{rosolowsky_structural_2008, goodman_role_2009, smullen_highly_2020, offner_turbulence_2022}. We use the dendrogram package \textsc{astrodendro}\footnote{www.dendrograms.org} to identify structures in column density maps of the clouds. We create column density maps of the central 5 parsecs of each cloud in the $x-y$ plane just before the onset of sink formation. We use a minimum column density threshold of $\rm 2\times10^{22} \, cm^{-2}$ and a minimum column density contrast of $\rm 8.3 \times 10^{21} \, cm^{-2}$. These correspond to the average column density of the cloud's initial conditions and the average column density of a $\rm 10^4 \, cm^{-3}$ core respectively. We require structures to contain a minimum of 10 pixels, corresponding to an area of $\rm 0.00026 \, pc^{2}$. Individual objects in the dendrograms are not directly studied as these are often transitory, however aggregate properties (e.g. the CMF) are generally time-invariant in dendrogram analysis \citep{smullen_highly_2020}, allowing them to be used here.

\begin{figure*}
    \centering
    \includegraphics[width=\textwidth]{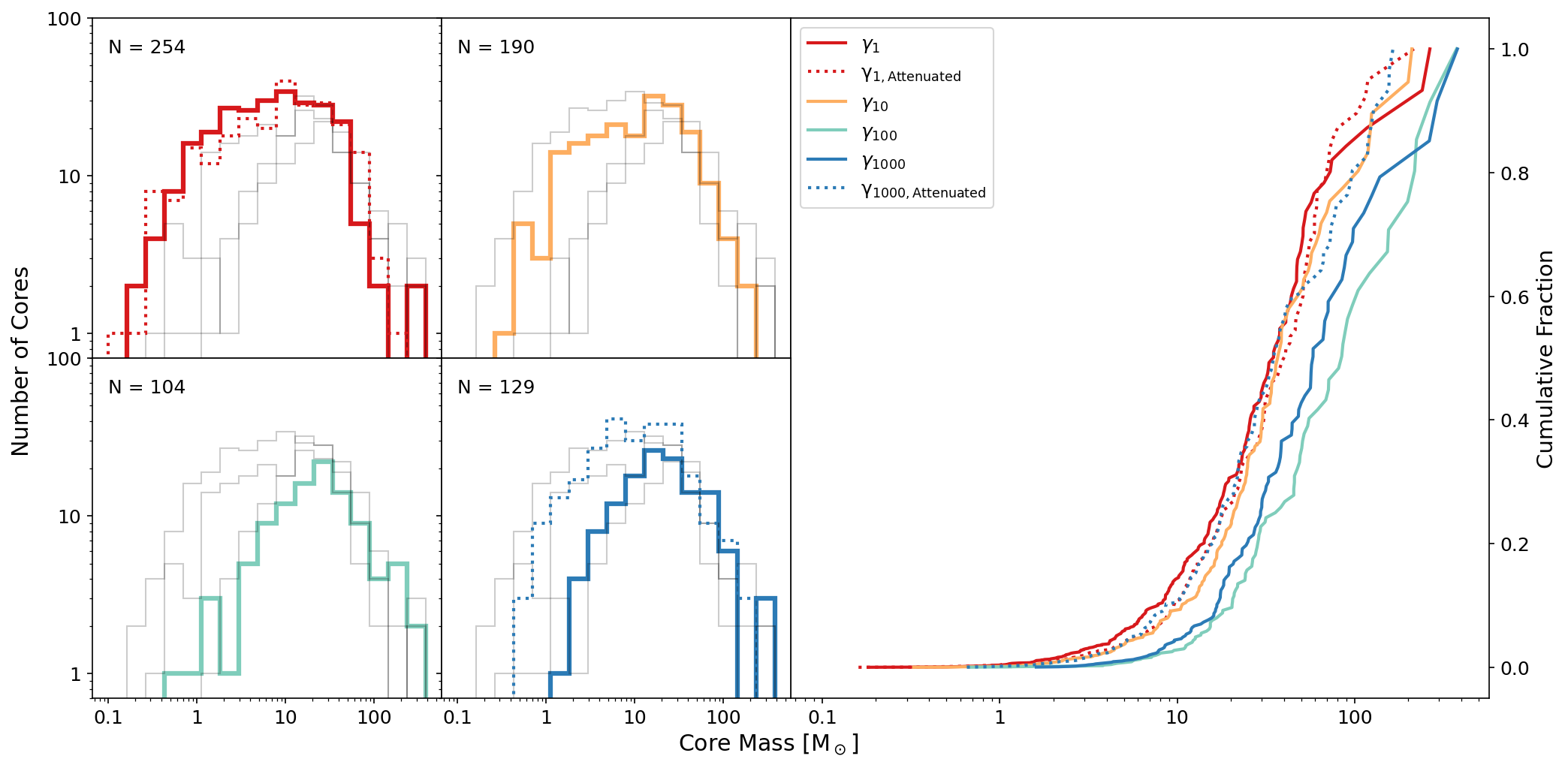}
    \caption{The mass distribution of cores just before the onset of sink formation in each cloud. Left hand panels show histograms of the core masses with the number of cores identified in the top-left. The right hand panel shows the cumulative distribution of core masses. Results from the simulations with cosmic ray attenuation are indicated with dotted lines.}
    \label{fig:coreMassFunction}
\end{figure*}

Figure \ref{fig:coreMassFunction} shows the mass function of the cores identified via dendrogram analysis. The peak of the core mass distribution shifts to higher masses with increasing $\rm \gamma_{SFR}$ and the distribution as a whole narrows. Small, sub-$\rm 1 \, M_\odot$, cores do not form in the high-$\rm \gamma_{SFR}$ clouds and instead a greater fraction of the cores are above $\rm 10 \, M_\odot$. The CMFs of the high-$\rm \gamma_{SFR}$ clouds are thus both top-heavy and bottom-light, as with the SMF. Thus, increasing $\rm \gamma_{SFR}$ suppresses the formation of cores and leads to a population of fewer, but more massive, cores overall. Observations of a starburst in the galactic bar (a high-$\rm \gamma_{SFR}$ environment) have noted a similar impact on the CMF \citep{motte_unexpectedly_2018}. The cumulative distribution reveals that the maximum core mass is different between the clouds, increasing with $\rm \gamma_{SFR}$. The minimum core mass also increases with $\rm \gamma_{SFR}$. Unlike the SMF, changing $\rm \gamma_{SFR}$ mainly acts to shift the distribution to higher masses and cut off the low-mass end, rather than fundamentally changing the shape of the distribution. 

The abundance of massive cores in the high-$\rm \gamma_{SFR}$ clouds may explain the distribution of sinks. Larger cores can fragment more than smaller ones, and produce richer groupings of sinks. The relative lack of massive cores in the low-$\rm \gamma_{SFR}$ clouds and the abundance of them in $\rm \gamma_{SFR}$ clouds is consistent with the average number of sinks produced by each core increasing with $\rm \gamma_{SFR}$. This result appears to contradict the finding of a top-heavy sink mass function, as a core that fragments more will produce more sinks that will competitively accrete with one another. Yet we have also shown that competitive accretion is weakened in these clouds. Therefore, despite producing more sinks per core, the cores must provide each sink with a larger reservoir of gas from which to accrete before competing with nearby sinks. The top-heavy core mass function indicates that this is the case.

The highest $\rm \gamma_{SFR}$ cloud presents an outlier due to being slightly less top-heavy than $\gamma_{100}$. This effect is present in the core mass function of both turbulent seeds, suggesting that it is not simply due to randomness. The highest-$\rm \gamma_{SFR}$ clouds experience compression due to the dense envelope we noted in Figure \ref{fig:wholeCloudOnsetSFR}. This compression may accelerate the collapse of the cloud, triggering more cores to form. The reduction in size of the cloud and more chaotic environment this produces may, however, suppress the formation of high-mass cores and lead to the change in the CMF we see.

\section{Cosmic Ray Attenuation} \label{sec:crAttenuation}

The inclusion of cosmic ray attenuation has a limited impact on the fiducial cloud. Figure \ref{fig:tempDensityEnd} shows that for this cloud, temperatures are only moderately reduced with attenuation and therefore the resulting core and sink mass functions are largely unchanged. This leads us to the conclusion that cosmic ray attenuation may not be dynamically significant in solar-type clouds with typical cosmic ray ionisation rates.

Attenuation, however, does have an effect on the high-$\rm \gamma_{SFR}$ clouds. Gas temperatures for number densities $\rm > 10^4 \, cm^{-3}$ are reduced by a similar amount to a reduction in magnitude of $\rm \gamma_{SFR}$. However, temperatures are largely unchanged for very low ($\rm < 10^4 \, cm^{-3}$) or very high ($\rm > 10^8 \, cm^{-3}$) number densities. This suggests that cosmic ray heating is important for setting the temperature at intermediate densities, but not so for very high or very low densities. In the high-$\rm \gamma_{SFR}$ clouds, we have already discussed how cosmic ray heating becomes more influential than photoelectric heating - explaining why the change to cosmic ray heating has a strong effect on the gas temperature. At high temperatures the gas temperature tends to that of the dust temperature, due to frequent gas-grain collisions that efficiently exchange energy between the two. The dust effectively acts as a thermostat with its temperature regulated by the ISRF rather than the CRIR.

The reduction in temperature in $\rm \gamma_{1000}$ resulting from the attenuation leads to a change in the core mass function. The CMF is no longer bottom-light, though it remains slightly top-heavy compared to $\rm \gamma_1$. This supports the notion that the number densities where the temperatures have been reduced the most ($\rm 10^4-10^6 \, cm^{-3}$) are critical to the formation of cores, and the thermal state here strongly influences the properties of cores. This relationship may be more complex than a simple dependence on temperature, however, as the CMF of the attenuated cloud does not resemble that of $\rm \gamma_{100}$ despite their similar temperatures. 

The sink mass function, on the other hand, follows a distribution with summary statistics between those of $\rm \gamma_{100}$ and $\rm \gamma_{1000}$. It is also top-heavy, and tracks the distribution of $\rm \gamma_{100}$'s sinks closely at high masses (but less so at low masses). This indicates that the sink mass function may not be as strongly influenced by attenuation as the core mass function, and that the change in the core mass function has not translated into a similar change in that of the sink particles. The accretion rate of the sink particles may be more influenced by the thermal state of the gas at very high densities, where the temperature is regulated by gas-grain interactions rather than cosmic ray heating.

These results suggest that cosmic ray attenuation may not be critical for simulating solar-neighbourhood like environments. It does not produce significant differences in the thermal state of the gas or the corresponding mass functions. Attenuation may however be important for high-CRIR environments. It can significantly impact the core mass function, and mildly affect the sink mass function and thermal state of the gas. 

\section{Discussion} \label{sec:discussion}

The results in the preceding sections suggest that increasing $\rm \gamma_{SFR}$ has a measurable impact on the fragmentation behaviour of molecular clouds. This change is reflected in the resulting mass functions and spatial distributions of the cores and sink particles. The results of this study imply that clouds in high-$\rm \gamma_{SFR}$ environments fragment into fewer (but generally more massive) cores that themselves typically fragment into larger groups of sink particles. These sinks accrete more rapidly and less competitively from the cores, resulting in a top-heavy mass function. The link between these changes and the thermodynamic state of the cloud is complex, and we will discuss how it changes the core and sink mass functions in turn. 

\subsection{Collapse into Cores}

The thermodynamics of a cloud impacts its ability to fragment in two ways. Firstly, a hotter cloud is more thermally supported against gravitational collapse. Increasing the temperature of a gas increases its Jeans mass, which is the preferred scale of fragmentation. Secondly, a warmer gas has a larger sound speed. The ability of turbulence to generate structure through shocks is dependent on the Mach number of those shocks, which is reduced in gasses with a large sound speed. These effects combine to drastically reduce the amount of structure that can form in the high-$\rm \gamma_{SFR}$ clouds. This is reflected in the change to the CMF, where we identify fewer cores overall, and those that we do find are generally more massive. The increased stability of the gas prevents the formation of low-mass cores, but massive cores are able to overcome this support. 

Since low mass core formation is suppressed, the peak of the CMF increases with $\rm \gamma_{SFR}$. This peak tracks well the Jeans mass at a number density of $\rm 10^6 \, cm^{-3}$ in each cloud, indicating that this density regime is important to core formation. For the runs with cosmic ray attenuation, the change in temperature at these densities produces a substantial change in the core mass function - further indicating that core formation is regulated by the thermal state of the cloud at such densities. This idea is supported by observations of pure Jeans mass fragmentation on the core scale \citep{palau_gravity_2015}. Simulations indicate that fragmentation is linked to changes in the effective equation-of-state of the gas, where fragmentation continues as long as the Jeans mass decreases sufficiently rapidly with increasing density \citep{jappsen_stellar_2005}. The density regime that we suggest cores form within is roughly isothermal and we suggest that this causes the fragmentation of the cloud into cores.

We do expect to see a shift of the CMF to larger masses in high-$\rm \gamma_{SFR}$ environments. \citet{motte_unexpectedly_2018} found a top-heavy core mass function in a Galactic Centre cloud undergoing a starburst - the exact kind of environment we have attempted to model here. Additionally, the theory presented by \citet{hennebelle_analytical_2008}, \citet{hennebelle_analytical_2009} and \citet{hopkins_stellar_2012} predicts that the CMF and SMF are regulated by the effective Mach number of a cloud. Their assertion that these mass functions will shift to lower masses as the Mach number increases agrees with the finding of top-heavy mass functions in the high-$\rm \gamma_{SFR}$ clouds (which have the lowest effective Mach number). \citet{gong_dense_2011} similarly suggests that lower Mach numbers will reduce and delay core formation.

We have thus far only considered the thermal physics of the gas as the primary driver of fragmentation and structure in the clouds as this is the focus of this study. All the simulations begin virialised, thus the initial amount of kinetic energy is the same for all the clouds. Cursory analysis reveals that the total kinetic energy present in the clouds increases over time, with a larger increase for higher values of $\rm \gamma_{SFR}$. Given that the turbulence is not driven, this additional energy must come from changes in the gravitational potential and thermal balance of the cloud that are influenced by $\rm \gamma_{SFR}$. We do not, however, find any changes in the virial balance of the cloud that would indicate that this change in kinetic energy is changing the ability of bound cores to form (see Figure \ref{fig:virialParams}). As a result of this, and because the different turbulent seeds produce similar changes in the CMF, we attribute the changes in structure in the clouds solely to changes in $\rm \gamma_{SFR}$.

\subsection{Fragmentation into Sinks}

Following their formation, cores gravitationally collapse and fragment. The Jeans mass decreases rapidly as the density of the core increases while its temperature decreases. This makes the cores highly Jeans-unstable and results in the production of clusters of sink particles. The amount of fragmentation is again dependent on the effective equation-of-state (or slope on the temperature-density diagram) of the gas, and a steeper slope incites more fragmentation \citep{li_effects_2003, jappsen_stellar_2005}. The steeper slope during collapse in the high-$\rm \gamma_{SFR}$ clouds explains why the cores in these clouds produce larger groups of sink particles. They are already more massive and the Jeans mass will decrease further as they collapse, than in the low-$\rm \gamma_{SFR}$ clouds, encouraging fragmentation. This fragmentation will continue unabated until the temperatures begin to increase with density at the opacity limit. We insert the sink particles before this occurs, and thus we may miss some of the fragmentation that may occur between the insertion of the sink particles and the formation of optically-thick cores. 

The fragmentation of cores into larger groups of sinks does not result in a more competitive environment for accretion, as the high-$\rm \gamma_{SFR}$ clouds instead produce top-heavy mass functions. The initial fragments in a core act as seeds from which the sinks then grow via accretion that is regulated by its environment \citep{maschberger_relation_2014, ballesteros-paredes_bondi-hoyle-littleton_2015, clark_emergent_2021}. The accretion onto the high-$\rm \gamma_{SFR}$ sinks is enhanced, and shows weaker evidence of competitive accretion than the fiducial clouds. The only way to reconcile the increased core fragmentation with a top-heavy sink mass function is for those cores to be more massive. Indeed, we have already shown that the cores in the high-$\rm \gamma_{SFR}$ clouds are generally more massive. This provides the sink particles with a richer accretion reservoir, and they can grow more massive before needing to compete with one another. Fragmentation may also be suppressed at very high densities thanks to the slightly increased temperatures in the high-$\rm \gamma_{SFR}$ clouds. If the core mass increases more rapidly with $\rm \gamma_{SFR}$ than the amount of fragmentation does, the sinks will each have more mass to accrete, leading to a top-heavy mass function. The results of the attenuated runs suggest that the thermal state of the gas at high densities may indeed be important for regulating sink growth, as the high-$\rm \gamma_{SFR}$ CMF is less bottom-light but the SMF is still top-heavy.

\subsection{Wider Context}

The most similar study to that presented here is that of \citet{guszejnov_effects_2022}, where ISRF values up to 100x solar were considered. They found an increase in stellar masses with the ISRF and a shallowing of the high-mass tail of the IMF. They noted higher sink formation rates and higher temperatures, especially at high densities. The change in the IMF they note is weak in comparison to that we find and their result of an increased star formation rate is in contrast to our own findings of a reduction of star formation rate with increasing $\rm \gamma_{SFR}$. However, \citet{guszejnov_effects_2022} did not vary the CRIR, which is an important heat source in the clouds we simulate (Figure \ref{fig:heatCoolFractional}). This may explain the discrepancy in gas temperatures and sink formation rates between our simulations. Still, the agreement on a change in the IMF/SMF to larger masses is encouraging, and both our study and theirs explains this effect similarly. On the other hand, \citet{chabrier_variations_2014} have suggested a link between Mach number and the IMF: increased Mach numbers may shift the IMF to lower masses. This is, however, disputed by \citet{bertelli_motta_imf_2016} who find the reverse. Our results tend to agree with \citet{chabrier_variations_2014}, as the core and sink mass functions become increasingly top-heavy as the effective Mach number decreases. Since the turbulence is not driven and can freely decay, the resistance to collapse provided by high Mach numbers is not present in the simulations and this may explain the discrepancy with the results of \citet{bertelli_motta_imf_2016}. \citet{whitworth_minimum_2024} recently studied the effect of changing the intensity of the ambient background radiation on the minimum mass for star formation, finding an increase in this minimum mass as the amount of radiation increases. As a result of this they find bottom-light mass functions, in agreement with our results for the SMF. Finally, \citet{klessen_stellar_2007} found a top-heavy IMF when modelling Galactic Centre environments and surmised that collapse sets in at densities of $\rm 10^5 \, cm^{-3}$ in agreement with our arguments about the CMF.

Comparison with \citet{guszejnov_effects_2022} reveals how critical cosmic rays are to the thermodynamics and fragmentation of the ISM. We saw in Figure \ref{fig:heatCoolFractional} that cosmic ray heating plays a pivotal role in the thermal balance of the ISM over a wide range of densities, helping to set the temperature of the gas where fragmentation occurs. Our findings support the idea of cosmic rays as regulators of star formation, as presented by \citet{papadopoulos_initial_2013}, thus highlighting the importance of an accurate cosmic ray treatment in any simulation of the ISM and star formation (see e.g. \citet{clark_temperature_2013}). This includes cosmic ray attenuation, for which there is some evidence of and work towards modelling \citep{padovani_cosmic-ray_2018, padovani_cosmic_2022}, alongside efforts to model cosmic rays more accurately overall \citep{girichidis_spectrally_2020, krumholz_cosmic_2020, girichidis_spectrally_2022, krumholz_cosmic_2022, girichidis_spectrally_2024}. The results from our attenuated runs indicate that cosmic ray attenuation may have a limited effect in solar neighbourhood environments, as we see little change in the $\rm \gamma_1$ simulations when attenuation is included. Attenuation appears most influential in setting the CMF in simulations with $\rm \gamma_{1000}$, negating the effects of increasing $\rm \gamma_{SFR}$ on the formation of low mass cores. The impact of attenuation on the SMF is comparatively weak, indicating that cosmic rays may not be important for regulating accretion. More work is needed to understand the full effect of cosmic ray attenuation on star formation. 

%% RE: REVIEWERS COMMENTS
The starburst environments we are modelling here are often observed at high redshifts, where low metallicities are more common. While regions with ongoing star formation will be naturally enriched due to said star formation, they may still have metallicities below solar. While a full exploration of the effects of metallicity is beyond the scope of this work, it is important to note the effect it may have on the results presented here.

A reduction in metallicity results in a reduced abundance of carbon, oxygen and dust. These are the primary coolants in the ISM, and their reduction will result in a higher temperature. This compounds with the effects of increasing $\rm \gamma_{SFR}$ we have already shown. This is corroborated by various metallicity studies, that show a decrease in metallicity results in an increase in temperature due to to a reduction in cooling and shielding (e.g. \cite{bate_statistical_2019, glover_star_2012}).

In the case of no metallicity (i.e Pop III star formation), increasing the intensity of cosmic rays exerts a net cooling effect as it catalyses the formation of $\rm H_2$ which in turn cools the gas \citep{hummel_first_2016}. However, the formation of $\rm H_2$ on dust grains becomes the dominant formation mechanism above metallicities of $\rm 10^{-3} \, Z_\odot$ \citep{cazaux_molecular_2004}, and in such cases cosmic-ray induced $\rm H_2$ formation is relegated to secondary importance. Therefore while in the majority of environments cosmic rays are primarily sources of heating, in some cases the reverse is true.

\subsection{Limitations and Caveats}

There are, however, caveats inherent to this study. Most importantly, we do not resolve individual stars or protostellar discs and can only comment on the combined system mass function. The large sink sizes we use mean that we will have included some systems that could have been disrupted before forming a protostar. The fragmentation of protostellar discs is also an important process that can limit the growth of stars and birth new ones \citep{peters_limiting_2010} and is essential to producing an IMF accurate for both stars and brown dwarfs \citep{bate_stellar_2012}. However, disc fragmentation may be suppressed when the discs are heated (either by the star or its environment), potentially further supporting our results \citep{bate_importance_2009}.

Additional criticisms of this study are the lack of radiative feedback from stars and the omission of magnetic fields. Both of these have been shown to play a role in the fragmentation of molecular clouds, and the delivery of an accurate IMF. However, we would argue that much of the impact of these mechanisms is irrelevant to the conclusions we have made.

Radiative feedback can impact the fragmentation and resulting stellar masses of a simulation, generally reducing fragmentation and leading to a top-heavy mass function \citep{krumholz_radiation_2006, bate_importance_2009, hennebelle_secondary_2020, hennebelle_influence_2022}. However, feedback from protostars tends to have its strongest effect on the circumstellar disc, inhibiting its fragmentation \citep{bate_importance_2009}. As we do not model this regime in the simulations it is unlikely that the heating would be relevant to the fragmentation that we see. On larger scales, radiative feedback is key to dissipating molecular clouds and halting star formation. However, we terminate the simulations before one would expect this dispersal to occur \citep{zhou_gas_2024} and it is unclear whether feedback alone would be enough to disrupt a cloud on the scale we have simulated \citep{guszejnov_starforge_2021}. The photoionisation shells and stellar winds we expect the larger (> 8 Msun) stars to produce can also impact fragmentation and stellar masses \citep{dale_early_2015}, but this effect is generally found to be weak due to the higher density of gas in star-forming cores compared to the wider cloud \citep{dale_massive_2013}, and the porous nature of clouds that allows much of the ionising radiation to escape \citep{dale_ionizing_2012, dale_before_2014}. The escaping radiation is part of what makes up the radiative feedback present in the external environment that we are modelling by increasing $\rm \gamma_{SFR}$.

On the other hand, magnetic fields play a nuanced role in the fragmentation of molecular clouds and subsequent star formation. In simulations they typically provide support against collapse that can reduce the amount of fragmentation seen \citep{lee_analytical_2017, ntormousi_core_2019, hennebelle_role_2019}. However, it is not clear how important this effect is. While magnetic fields change the dynamics and evolution of the ISM, they do not drastically reduce star formation rates and both turbulence and gravity can dominate over magnetic fields in different scenarios \citep{hennebelle_role_2019, hennebelle_influence_2022}. It also is not accepted how the effects of magnetic fields manifest in the IMF, as magnetic fields have also been shown to increase the amount of fragmentation into low mass stars \citep{li_lowering_2010}. While magnetic fields and radiative feedback, taken together, are required to produce an accurate IMF, they are not required to determine the peak and slope of the IMF (of interest to this study) \citep{hennebelle_influence_2022}. Overall, the effects of magnetic fields are complex and not the focus of this work. It is unclear how they might interact with $\rm \gamma_{SFR}$ to change the results, or if at all. 

Overall, while inclusion of these mechanisms is crucial to realistically simulating molecular clouds, we do not expect that doing so would change the conclusions we have drawn.

\section{Conclusions} \label{sec:conclusion}

In this work we have explored the behaviour of molecular clouds in starburst environments. We have investigated their thermodynamic properties, fragmentation behaviour and the subsequent sink particles that they form. The conclusions of this study can be summarised as follows:

i) Increasing the ISRF \& CRIR heats the gas in molecular clouds, increasing the equilibrium temperature at all densities. This causes an extension of the WNM to higher densities, disruption to the CNM-molecular transition and a delay to dust-gas coupling. The change in temperature increases the average Jeans mass of the clouds and decreases the effective Mach number of turbulent shocks.

ii) The change in the thermodynamic state of the gas leads to a change in the structure of the clouds on all scales. Structures become both larger in mass and fewer in number. This shifts the core mass function to higher masses and we find an increasingly top-heavy and bottom-light CMF with increasing $\rm \gamma_{SFR}$.

iii) Cores in high-$\rm \gamma_{SFR}$ clouds fragment into larger groups of sink particles as they collapse, leading to a change to the spatial distribution of sinks. Sinks in the high-$\rm \gamma_{SFR}$ clouds are members of rich clusters that accrete from massive cores.

iv) Increasing $\rm \gamma_{SFR}$ produces a top-heavy system mass function that follows a shallow high-mass tail. Due to the larger core masses, sinks accrete more rapidly and do so less competitively, contrary to expectations. We suggest that core masses are large enough in the high-$\rm \gamma_{SFR}$ clouds to facilitate both increased fragmentation and enhanced accretion. The reduction in the number of cores that form, however, leads to sink formation being delayed and reduced overall in these clouds.

v) Cosmic ray attenuation does not have a strong effect in solar-neighbourhood environments. Instead its greatest effect is on the core mass function of high-$\rm \gamma_{SFR}$ clouds, reducing the suppression of low mass core formation. The sink mass function and gas temperatures are only weakly affected compared to the CMF.

These conclusions lead us to paint a picture where clouds in high ISRF \& CRIR environments produce fewer, but more massive, cores that subsequently fragment into richer clusters of sinks that accrete at an enhanced rate. The size of the sink particles prevents us from drawing conclusions about the IMF in these environments, but we can assert that both the CMF and SMF are top-heavy. The results of this work imply that star formation in high SFR environments differs from that of the solar neighbourhood and that these differences must be taken into account when considering such environments.

\section*{Acknowledgements}

This work used the Isambard 2 UK National Tier-2 HPC Service (\url{http://gw4.ac.uk/isambard/}) operated by GW4 and the UK Met Office, and funded by EPSRC (EP/T022078/1). This research also used the supercomputing facilities at Cardiff University operated by Advanced Research Computing at Cardiff (ARCCA) on behalf of the Cardiff Supercomputing Facility and the HPC Wales and Supercomputing Wales (SCW) projects. We acknowledge the support of the latter, which is part-funded by the European Regional Development Fund (ERDF) via the Welsh Government.

We thank the referee for an encouraging and insightful report that improved the quality of this manuscript.

FDP and PCC acknowledge the support of a consolidated grant (ST/W000830/1) from the UK Science and Technology Facilities Council (STFC).

RSK and SCOG acknowledge support from the European Research Council via the ERC Synergy Grant “ECOGAL” (project ID 855130), from the German Excellence Strategy via the Heidelberg Cluster of Excellence (EXC 2181 - 390900948) “STRUCTURES”, and from the German Ministry for Economic Affairs and Climate Action in project “MAINN” (funding ID 50OO2206). They also ackowledge local computing resources provided by the Ministry of Science, Research and the Arts (MWK) of {\em The L\"{a}nd} through bwHPC and the German Science Foundation (DFG) through grant INST 35/1134-1 FUGG and 35/1597-1 FUGG, and also data storage at SDS@hd funded through grants INST 35/1314-1 FUGG and INST 35/1503-1 FUGG. RSK also thanks the 2024/25 Class of Radcliffe Fellows for company and for highly interesting and stimulating discussions.

PG acknowledges support from the European Research Council via the ERC Synergy Grant “ECOGAL” (project ID 855130).

%%%%%%%%%%%%%%%%%%%%%%%%%%%%%%%%%%%%%%%%%%%%%%%%%%

\section*{Data Availability}

The simulation snapshot data is available upon request to MTC. The public version of \textsc{arepo} is freely available at \url{github.com/dnelson86/arepo}. Codes used for analysis and generation of Figures are available at \url{github.com/clonematthew/stardustCodes}.

%%%%%%%%%%%%%%%%%%%% REFERENCES %%%%%%%%%%%%%%%%%%

\bibliographystyle{mnras}
\bibliography{references}

%%%%%%%%%%%%%%%%%%%%%%%%%%%%%%%%%%%%%%%%%%%%%%%%%%

%%%%%%%%%%%%%%%%% APPENDICES %%%%%%%%%%%%%%%%%%%%%

\appendix

\section{Initial Chemical Species Abundances}

Table \ref{tab:initSpecies} shows the initial chemical species abundances for each of the simulations. The equilibrium abundance of each of the tracked species in the chemical network changes with the ISRF and CRIR, as many of the rates depend in some way on these variables. Therefore we run low-resolution simulations of our clouds until the chemical abundances at the initial number density ($\rm 10^3 \, cm^{-3}$) are stable. The test simulations are initialised with fully atomic hydrogen and carbon. The relative abundances of carbon, oxygen and metals relative to hydrogen is the same for all the clouds. 

\begin{table*}
    \centering
    \caption{The initial abundances of all chemical species for each simulation run. These abundance values were determined by running a low-resolution simulation version of each setup until chemical equilibrium was reached. The abundance of each species at a number density of $10^3 \:\rm cm^{-3}$ was then chosen as the initial abundance for the simulations. The total C and O abundances relative to hydrogen were kept to $x_{\rm C} = 1.4 \times 10^{-4}$ and $x_{\rm O} = 3.2 \times 10^{-4}$ respectively, in accordance with \citet{sembach_modeling_2000}.} 
    
    \begin{tabular}{c|c|c|c|c|c|c|c|c|c}
        \hline
        Simulation & $\rm x_{H_2}$ & $\rm x_{H+}$ & $\rm x_{C+}$ & $\rm x_{CO}$ & $\rm x_{CHx}$ & $\rm x_{OHx}$ & $\rm x_{HCO+}$ & $\rm x_{He+}$ & $\rm x_{M+}$ \\ 
        \hline\hline
        $\gamma_1$    & 0.358 & $9.93 \times 10^{-7}$ & $3.95 \times 10^{-6}$ & $3.64 \times 10^{-5}$ & $8.39 \times 10^{-10}$ & $1.94 \times 10^{-7}$ & $5.68 \times 10^{-10}$ & $1.46 \times 10^{-7}$ & $1.05 \times 10^{-5}$ \\
        $\gamma_{10}$    & 0.363 & $1.04 \times 10^{-5}$ & $2.67 \times 10^{-5}$ & $1.06 \times 10^{-5}$ & $2.48 \times 10^{-9}$  & $2.81 \times 10^{-7}$ & $5.12 \times 10^{-10}$ & $1.51 \times 10^{-6}$ & $1.36 \times 10^{-5}$ \\
        $\gamma_{100}$   & 0.276 & $1.40 \times 10^{-4}$ & $6.99 \times 10^{-5}$ & $1.57 \times 10^{-5}$ & $1.25 \times 10^{-9}$  & $1.49 \times 10^{-6}$ & $2.52 \times 10^{-9}$  & $1.24 \times 10^{-5}$ & $1.48 \times 10^{-5}$ \\
        $\gamma_{1000}$  & 0.034 & $4.06 \times 10^{-3}$ & $1.22 \times 10^{-4}$ & $3.19 \times 10^{-7}$ & $7.48 \times 10^{-11}$ & $2.71 \times 10^{-7}$ & $1.21 \times 10^{-10}$ & $4.10 \times 10^{-4}$ & $1.50 \times 10^{-5}$ \\
        \hline
    \end{tabular}
    \label{tab:initSpecies}
\end{table*}

\section{Jeans Masses}

In this work we use the Jeans mass \citep{jeans_stability_1902}, the minimum mass that a uniform density sphere of gas of a given temperature and density needs to overcome thermal support and gravitationally collapse, for analysis and comparison. We use the following to calculate this quantity;

\begin{equation}
    M_{\rm Jeans} = \left(\frac{\pi k_B}{G}\right)^{3/2} \left(\frac{X_{H} T^3}{n \mu^3 m_p^4}\right)^{1/2}
\end{equation}
where $k_B$ is the Boltzmann constant, G is the gravitational constant, T is the temperature of the gas, n the number density of the gas, $\rm \mu$ the mean molecular weight of the gas in atomic mass units, $m_p$ the mass of a proton, and $X_{H}$ is the fractional abundance of hydrogen by mass.

\section{Cosmic Ray Attenuation} \label{sec:crAttenuationModel}

\begin{figure}
    \centering
    \includegraphics[width=\columnwidth]{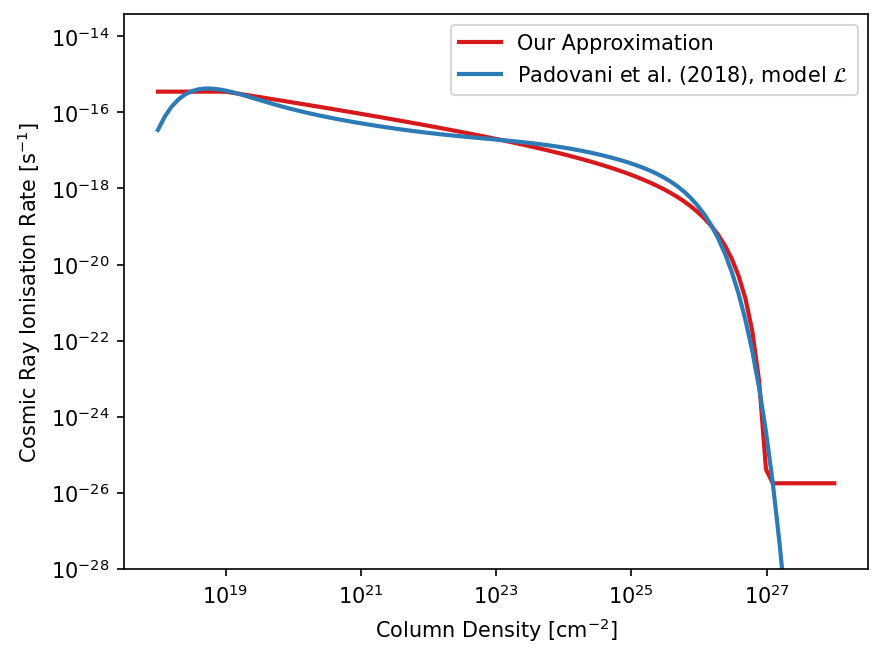}
    \caption{The cosmic ray ionisation rate against column density using both our approximation and the model by \citet{padovani_cosmic-ray_2018} for the solar neighbourhood value of $\rm 3.5 \times 10^{-16} \, s^{-1}$.}
    \label{fig:crAttenuation}
\end{figure}

We replicate runs $\rm \gamma_{1}$ and $\rm \gamma_{1000}$ including a simple approximation for cosmic-ray attenuation based on the parameterisation by \citet{padovani_cosmic-ray_2018}. We adopt a simple column-density based approach where the cosmic ray ionisation rate decreases with increasing column density, 

\begin{equation}
    \xi(\rm N_{H}) =
    \begin{cases}
    \xi_0, & \text{for $\rm N_H \leq 10^{19} \, cm^{-2}$} \\ 
    \xi_0 \, \frac{10^{-10.48 - \frac{\log \rm{N_{H}}}{4} + \frac{2.13}{\log \rm{N_{H}} - 27.25}}}{3.25\times 10^{-16}}, & \text{for $\rm 10^{19} < N_H < 10^{27} \, cm^{-2}$} \\
    1.78\times10^{-26}, & \text{for $\rm N_H \geq 10^{27} \, cm^{-2}$}
    \label{eq:crAtt}
    \end{cases}
\end{equation}
where $\rm N_{H}$ is the total column density of hydrogen and $\rm \xi_0$ is the base cosmic ray ionisation rate quoted in Table \ref{tab:simulations}. Figure \ref{fig:crAttenuation} compares our parameterisation (Equation \ref{eq:crAtt}) with the model of \citet{padovani_cosmic-ray_2018}. We adopt a constant value of the cosmic ray ionisation rate at very high column densities for numerical convenience. By a column density of $\rm 10^{27} \, cm^{-2}$ the attenuated rate is low enough to have a negligible impact on the chemistry and thermodynamics of the gas, such that any additional reduction would be superfluous and computationally expensive.

\section{Chemical Abundances}

\begin{figure*}
    \centering
    \includegraphics[width=\textwidth]{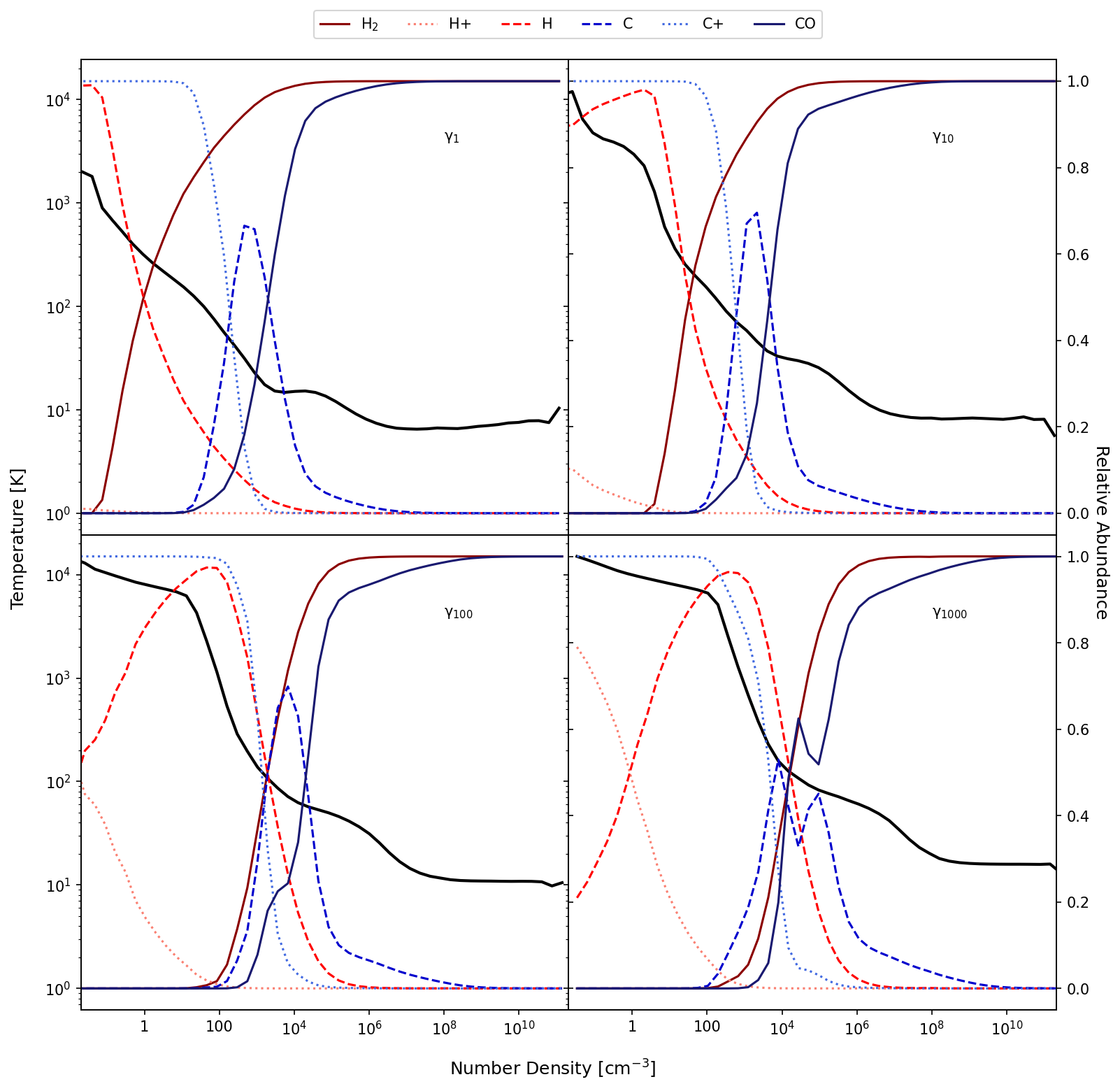}
    \caption{The relative abundances of species containing hydrogen and carbon across bins of number density for each cloud at the end of the simulation. The black line in each panel shows the temperature-density curve of each cloud. Abundances are relative to the maximum abundance of that species.}
    \label{fig:fracAbunds}
\end{figure*}

Figure \ref{fig:fracAbunds} shows the relative abundances of species containing hydrogen and carbon. Hydrogen species are shown in red, and carbon-containing species are shown in blue. Note that abundances are normalised relative to their maximum abundance, so the $\rm 1\times 10^{-4}$ relative abundance of carbon to hydrogen is not seen. We overplot the temperature-density curve of each cloud in black, showing the link between chemical and thermodynamic transitions.

\section{Raw Heating and Cooling Rates} \label{sec:ratesExtra}

\begin{figure*}
    \centering
    \includegraphics[width=\textwidth]{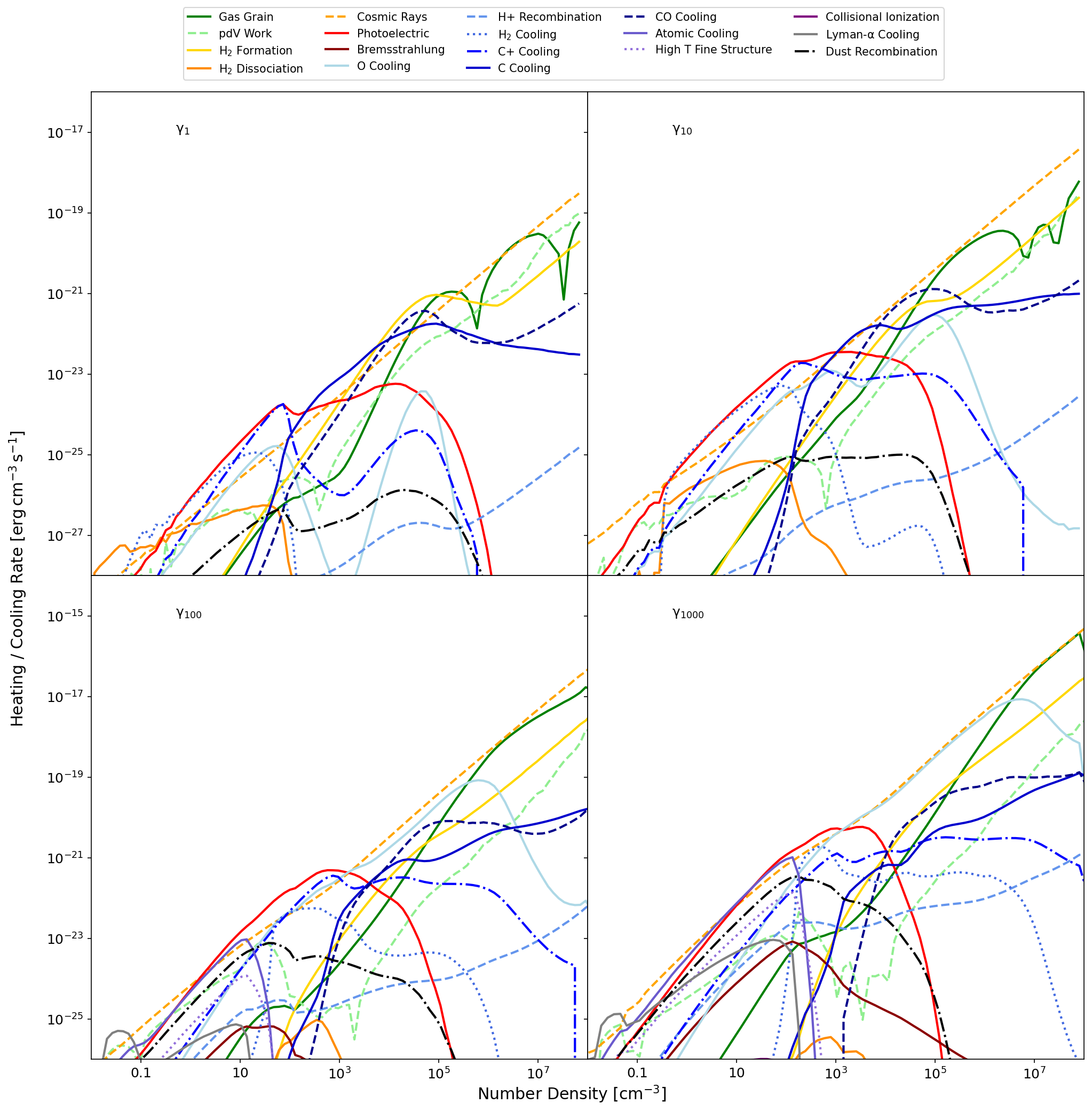}
    \caption{The heating and cooling rates per unit volume included in our chemical network and their raw rate against number density. Rates are binned into equal width bins of number density and a mass-weighted mean rate is found for each bin/rate. Rates are coloured depending on whether they heat (reds and oranges), cool (blues and purples) or can do either (greens).}
    \label{fig:heatCoolRatesRaw}
\end{figure*}

\begin{figure*}
    \centering
    \includegraphics[width=\textwidth]{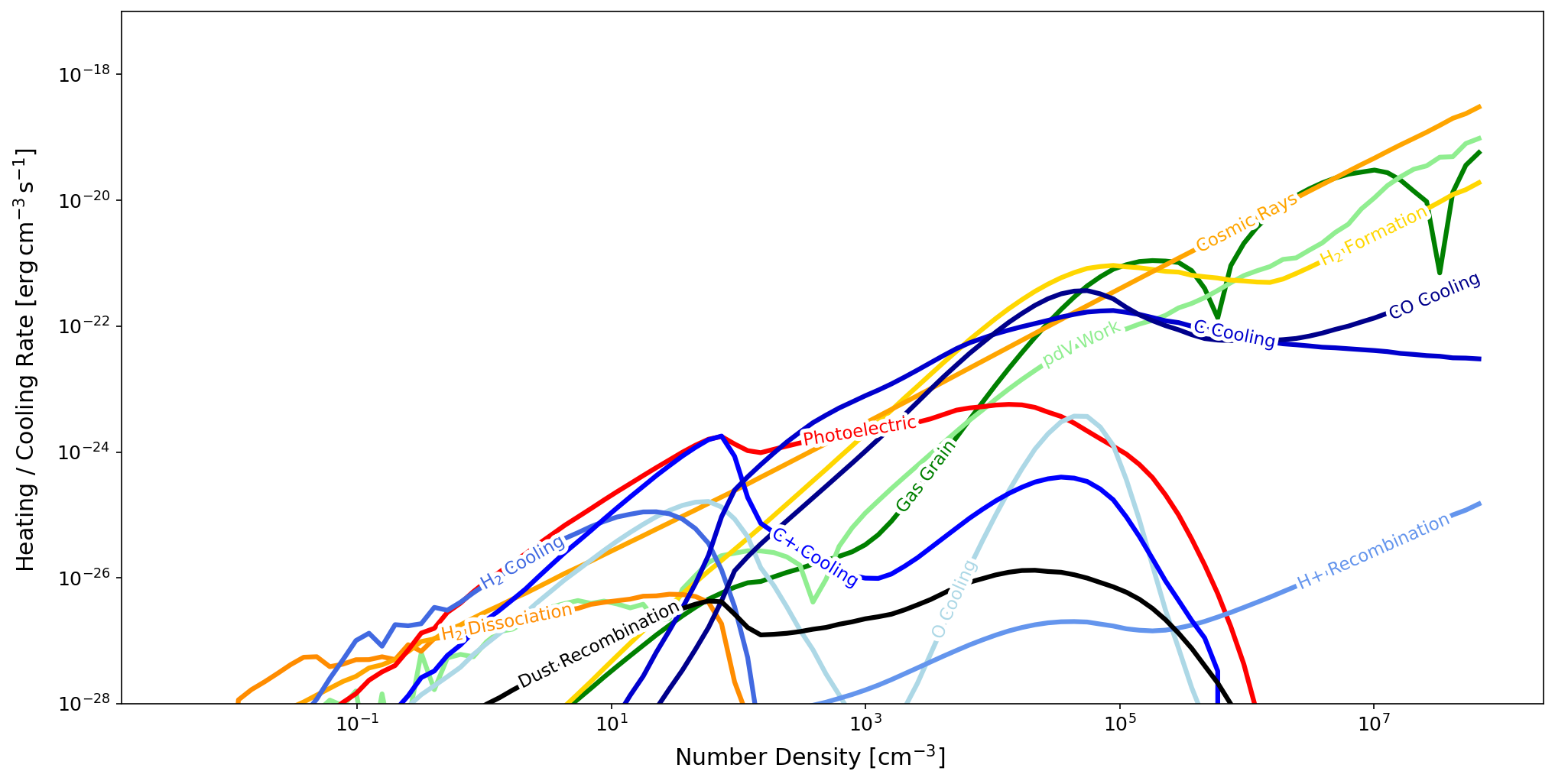}
    \caption{The heating and cooling rates included in our chemical network and their raw rate against number density for $\gamma_1$. Rates are binned into equal width bins of number density and a mass-weighted mean rate is found for each bin/rate. Rates are coloured depending on whether they heat (reds and oranges), cool (blues and purples) or can do either (greens).}
    \label{fig:uv1Rates}
\end{figure*}

Figure \ref{fig:heatCoolFractional} showed the fractional heating and cooling rates, normalised for each bin of number density. We do this as the value of many of the rates increases with density, obscuring the importance of rates at low densities. In Figures \ref{fig:heatCoolRatesRaw} and \ref{fig:uv1Rates} we instead plot the raw rate values.

\section{Attenuated Heating and Cooling Rates}

\begin{figure*}
    \centering
    \includegraphics[width=\textwidth]{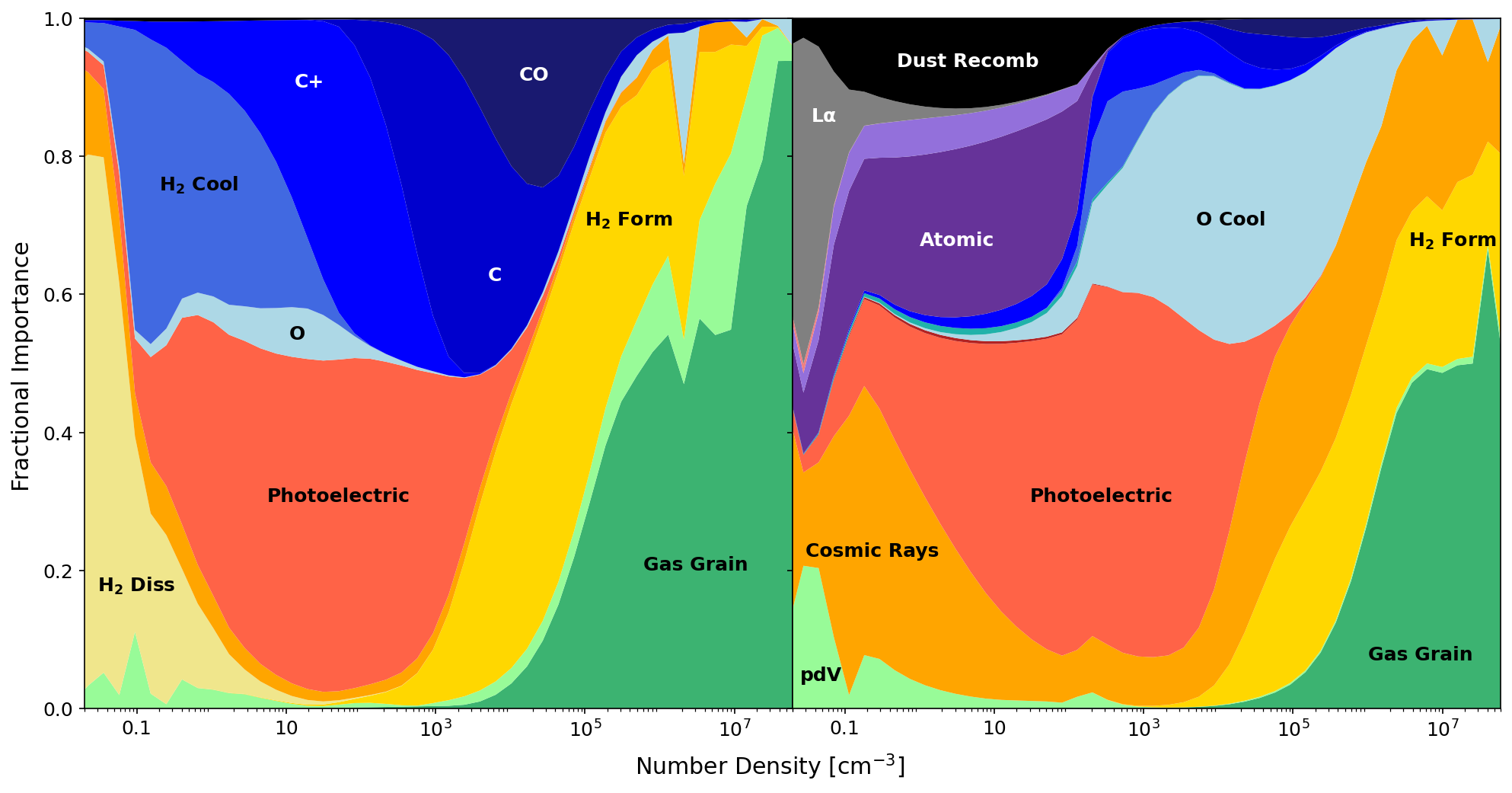}
    \caption{The heating and cooling rates included in our chemical network and their fractional importance against number density for the runs with cosmic ray attenuation. Rates are binned into equal width bins of number density and a mass-weighted average rate is found for each bin/rate. The rates are then normalised to account for the changing sum of raw rate values with number density. Rates are coloured depending on whether they heat (reds and oranges), cool (blues and purples) or can do either (greens). The left panel shows $\rm \gamma_1$ and the right panel $\rm \gamma_{1000}$.}
    \label{fig:heatCoolAtt}
\end{figure*}

Figure \ref{fig:heatCoolAtt} shows the fractional importance of the different heating and cooling rates in our network for the simulations with cosmic ray attenuation. The contribution from cosmic ray heating is reduced in both runs, almost completely in $\rm \gamma_1$ and by a significant amount in $\rm \gamma_{1000}$. At high densities, this reduction is compensated by an increase in the importance of gas grain and $\rm H_2$ formation heating. The gas temperature at high densities is largely independent of density despite this change (Figure \ref{fig:tempDensityEnd}), as it is regulated not by cosmic ray heating but by the dust temperature. The efficiency of dust emission ensures that any additional heating is radiated away, or any additional cooling is balanced by collisional heating. At lower densities, the gas and dust are not well coupled and changes in the cosmic ray heating rate can make a difference to the overall thermal state of the gas.

\section{Virial Balance} \label{sec:virial}

\begin{figure}
    \centering
    \includegraphics[width=\columnwidth]{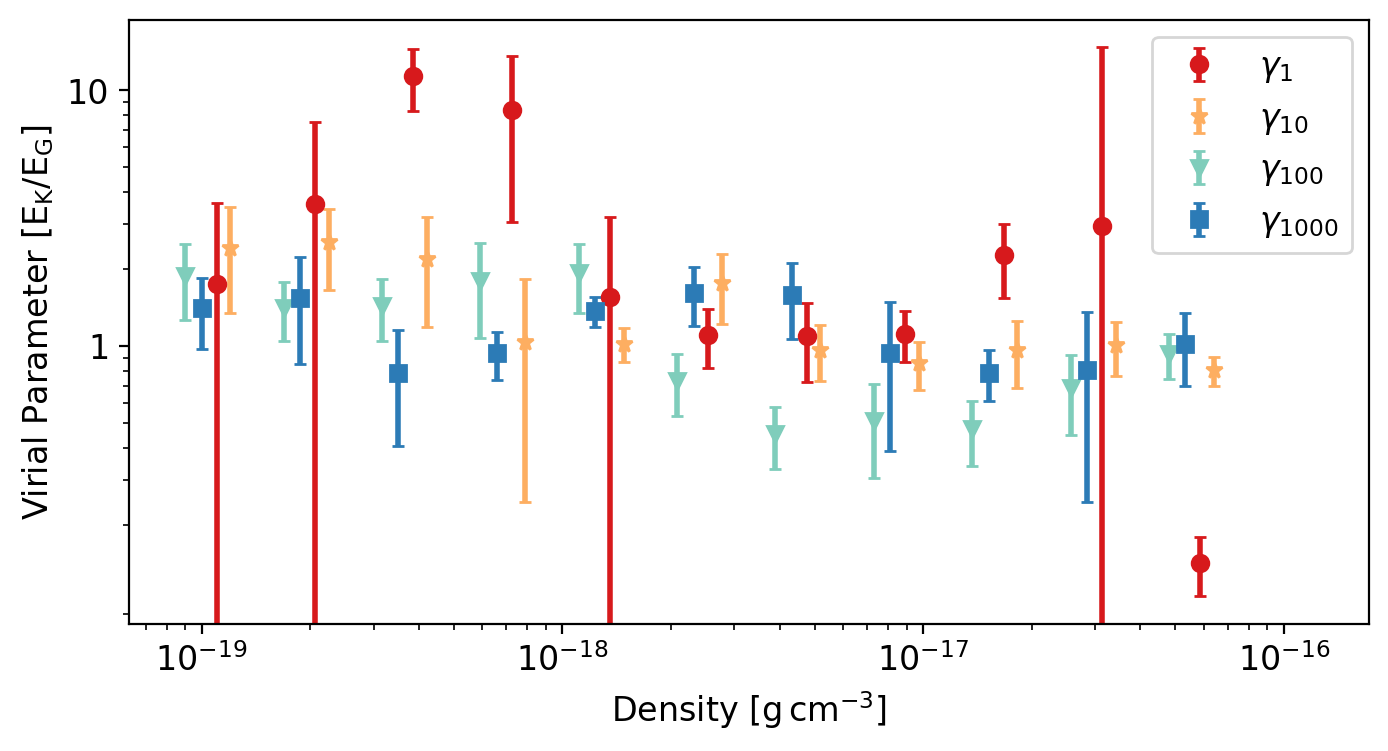}
    \caption{The average virial parameter against density for each simulation run. We randomly select 50 particles in each density bin and determine the virial parameter of the particles within a local Jeans length. We determine the mean and standard deviation from the virial parameters of the particles in each bin.}
    \label{fig:virialParams}
\end{figure}

The dynamics of the cloud are not the focus of this study. However, it is important to first establish that the fragmentation and collapse of the clouds is not strongly influenced by the changing kinetic energy of the clouds. In Figure \ref{fig:virialParams} we plot the virial parameter of a set of 50 randomly selected particles over a range of densities for each simulation, to establish if the virial balance of the cloud is changing. The Figure shows no systematic change in the average virial parameter with density, indicating that the dynamical states of the clouds are comparable to one another.

% Don't change these lines
\bsp	% typesetting comment
\label{lastpage}
\end{document}